\documentclass[final]{siamltex}
\usepackage{lmodern,eurosym,ae,tabularx,amssymb,url,epsf,epsfig,amsmath,amsfonts,graphicx,dsfont}
\font\dsrom=dsrom10 scaled 1200           
\newtheorem{thm}{Theorem}[section]
\newtheorem{prop}[thm]{Proposition}
\newtheorem{lemme}[thm]{Lemma}
\newtheorem{de}[thm]{Definition}
\newtheorem{rmq}[thm]{Remark}
\newenvironment{pf}{\ \\ {\it Proof of}}{\hfill\mbox{$\diamond$}\medskip}
\renewcommand{\P}{\mathbb{P}}
\newcommand{\indicatrice}{\textrm{\dsrom{1}}} 
\newcommand{\F}{{\mathcal F}}
\newcommand{\R}{{\mathbb R}}
\newcommand{\N}{{\mathbb N}}
\newcommand{\Z}{{\mathbb Z}}
\newcommand{\E}{{\mathbb{E}}}
\renewcommand{\P}{{\mathbb{P}}}
\newcommand\ind[1]{{\bf 1}_{#1}}

\title{Connecting discrete and continuous lookback or hindsight options in exponential L\'evy models}
\author{El Hadj Aly Dia\thanks{Universit\'e Paris-Est, Laboratoire d'Analyse et de Math\'ematiques Appliqu\'ees, UMR CNRS 8050, 5 bd. Descartes, Champs-sur-Marne, $77454$ Marne-la-Vall\'ee, France ({\tt dia.eha@gmail.com}).} \and Damien Lamberton\thanks{Universit\'e Paris-Est, Laboratoire d'Analyse et de Math\'ematiques Appliqu\'ees, UMR CNRS 8050, 5 bd. Descartes, Champs-sur-Marne, $77454$ Marne-la-Vall\'ee, France ({\tt damien.lamberton@univ-mlv.fr}).}}
\date{}
\begin{document}

\maketitle
  
\begin{abstract}
Motivated by the pricing of lookback options in exponential L\'evy models, we study the difference
between the continuous and discrete supremum of L\'evy processes. 
In particular, we extend the results of Broadie et al. (1999) to \emph{jump diffusion} models.
We also derive bounds for general exponential L\'evy models.
\end{abstract}
\pagestyle{myheadings}
\thispagestyle{plain}
\markboth{E. H. A. DIA AND D. LAMBERTON}{CONNECTING DISCRETE AND CONTINUOUS OPTIONS}
\section{Introduction}
\label{sec:intro}

The payoff of a lookback option typically depends on the maximum or the minimum of the underlying 
stock price. The maximum can be evaluated in continuous or discrete time depending on the contract.
In the Black-Scholes setting,  Broadie, Glasserman and Kou (1999 and 1997) derived a number of results relating discrete and continuous 
path-dependent options. In particular, they obtained continuity correction formulas for lookback, barrier
and hindsight options. The purpose of this paper is to establish similar results for exponential L\'evy
models. We will focus on lookback or hindsight options, leaving the treatment of barrier options
to another paper.

Our results are based on the analysis of the difference between the discrete and continuous maximum
of a L\'evy process. In the case of a L\'evy process with finite activity and a non zero Brownian part, we 
extend (see Theorem~\ref{supremum_error_law_convergence}) the theorem of  Asmussen, Glynn and Pitman (1995) which is the key to the continuity correction
formulas for lookback options in Broadie, Glasserman and Kou (1999). This allows us to extend
these formulas to jump-diffusion models. We also establish estimates for the $L_1$-norm
of the difference of the continuous and discrete maximum of a general L\'evy process.
These estimates are based on Spitzer's identity, which relates the expectation of the supremum 
of sums of iid random variables to a weighted sum of the expectations of the positive 
parts of the partial sums. In the case of L\'evy processes with finite activity, we derive an expansion
up to te order $o(1/n)$, where $n$ is the number of dates in the discrete supremum, 
see Theorem~\ref{weak_error}. In the case of infinite activity, we have precise upper bounds
(see Theorem~\ref{error}). We also derive an expansion in the case of L\'evy processes with finite 
variation (see Theorem~\ref{error2}).

The paper is organized as follows. In the next section, we recall some basic facts about  real L\'evy processes.  In section~\ref{sec:spitzer},  we state  Spitzer's identity for L\'evy Processes and use it to analyse the expectation of the difference of the continuous and discrete maximum
of a general L\'evy process. Section~\ref{sec:asmussentheorem} is devoted to the extension of the theorem of Asmussen et al. The last two sections are devoted to financial applications.
In Section~\ref{sec:contcorr}, we derive continuity  corrections for lookback options in \emph{jump-diffusion} models, and  in Section~\ref{sec:bounds}, we give upper bounds  
for the case of general exponential  L\'evy models.
\section{Preliminaries}
\label{sec:prelim}
A real L\'evy process $X$ is characterized by its generating triplet $(\gamma,\sigma^2,\nu)$, where $(\gamma,\sigma)\in\R\times\R^{+}$, and $\nu$ is a Radon measure on $\R\backslash\{0\}$ satisfying
\begin{eqnarray*}
  \int_{\R} \left(1\wedge x^2\right)\nu(dx)<\infty.
\end{eqnarray*} 
By the L\'evy-It\^o decomposition, $X$ can be written in the form
\begin{equation}\label{levy_Ito}
X_t=\gamma t + \sigma B_t +X_t^l + \lim_{\epsilon \downarrow 0} \widetilde{X}_t^{\epsilon},
\end{equation}
with
\begin{equation*}
X_t^l=\int_{|x| > 1,s\in[0,t]} xJ_X(dx \times ds)\equiv \sum_{0 \leq s \leq t}^{|\Delta X_s| \geq 1}\Delta X_s 
\end{equation*}
\begin{eqnarray*}
\widetilde{X}_t^{\epsilon}&=& \int_{\epsilon \leq |x| \leq 1,s\in[0,t]} x\widetilde{J}_X(dx \times ds)\equiv \sum_{0 \leq s \leq t}^{\epsilon \leq |\Delta X_s| < 1}\Delta X_s - t\int_{\epsilon \leq |x| \leq 1}x\nu(dx).
\end{eqnarray*}
Here $J$ is a Poisson measure on $\R \times [0,\infty)$ with intensity $\nu(dx)dt$, $\widetilde{J}_X(dx \times ds)=J_X(dx \times ds)-\nu(dx)ds$ and $B$ is a standard Brownian motion. We also
have the  L\'evy-Khinchine formula for the characteristic function of $X_t$. Namely
\begin{equation*}
\E e^{iuX_t}=e^{t\varphi(u)}, \quad u\in\R,
\end{equation*}
where $\varphi$ is given by
\begin{equation}\label{levy-kintchine}
\varphi(u)=i\gamma u-\frac{\sigma^2 u^2}{2}+\int_{\R} (e^{iux}-1-iux \indicatrice_{|x| \leq 1})\nu(dx).
\end{equation}
We say that  $X$ has finite activity if the L\'evy measure $\nu$ is finite ($\nu\left(\R\right)<\infty$). We then have
\begin{equation}\label{levyactivitefinie}
X_t=\gamma_0t+\sigma B_t + \sum_{i=1}^{N_t}Y_i,
\end{equation}
where $N$ is a Poisson process with rate $\lambda=\nu(\R)$, $\left(Y_i\right)_{i\geq1}$ are i.i.d. random variables
 with common distribution $\frac{\nu(dx)}{\nu(\R)}$ and
\begin{equation}\label{gamma}
\gamma_0=\gamma  - \int_{|x| \leq 1}x \nu(dx).
\end{equation}
This is a \emph{jump-diffusion} process. If the jump part of $X$ has finite variation (which is equivalent to $\int_{|x|\leq1}|x|\nu\left(dx\right)<\infty$), then
\begin{equation}\label{levyactiviteinfinievf}
X_t=\gamma_0t+\sigma B_t + \int_{x\in\R,s\in[0,t]} xJ_X(dx \times ds),
\end{equation}
with $\gamma_0$ given by \eqref{gamma}. Note that $X$ is a finite variation L\'evy process if and only if $\sigma=0$ and $\int_{|x|\leq1}|x|\nu\left(dx\right)<\infty$. Moreover, $X$ is integrable if only if $\int_{|x|>1}|x|\nu(dx)<\infty$.

\section{Spitzer's identity and applications}
\label{sec:spitzer}
In this section we will first state Spitzer's identity for L\'evy processes (we refer to \cite{asmussen87},
Proposition 4.5, p. 177
for the classical form of Spitzer's identity). Then we will use this result to derive expansions for the error between the continuous and discrete supremum of L\'evy processes.
\begin{de}\label{extremumdef}
We define
\begin{eqnarray*}
 M_t^{X}=\sup_{0\leq s\leq t}X_{s}, \ \ M_t^{X,n}=\max_{0\leq k\leq n}X_{\frac{kt}{n}}.
\end{eqnarray*}
\end{de}
When there is no ambiguity we can remove the super index $X$. 
\begin{rmq}\rm
 Note that $M_t$ is integrable for all $t>0$ if and only if
$\int_{x>1}x\nu(dx)$ is finite. We also have, for all $\alpha>0$, $\E e^{\alpha M_t}<\infty$ if only if 
$\int_{x>1}e^{\alpha x}\nu(dx)$ is finite.
\end{rmq}

In the setting of L\'evy processes, we have the following version of Spitzer's identity.
\begin{prop}\label{spitzer}
If $X$ is a L\'evy process with generating triplet $(\gamma,\sigma^2,\nu)$ satisfying $\int_{x> 1}x\nu(dx)<\infty$, then
\begin{eqnarray*}
\E M_t^n=\sum_{k=1}^n \frac{\E X_{k\frac{t}{n}}^{+}}{k}, \quad \E M_t=\int_0^t \frac{\E X_s^{+}}{s} ds. 
\end{eqnarray*} 
\end{prop}
For the proof of  the above result, we need some estimates for $\E M_t$ with respect to $t$.
\begin{prop}\label{levy_supremum_error}
Let $X$ a L\'evy process with generating triplet $(\gamma,\sigma^2,\nu)$ satisfying $\int_{x > 1} x \nu(dx) < \infty$, then
\begin{equation*}
\E M_t\leq\left(\gamma^{+}+\int_{x>1}x\nu(dx)\right)t + \left(\sigma \sqrt{\frac{2}{\pi}}+2\sqrt{\int_{|x|\leq1} x^2 \nu(dx)}\right)\sqrt{t}.
\end{equation*}
If in addition $\int_{|x|\leq1}|x|\nu(dx)<\infty$, then
\begin{equation*}
\E M_t\leq\left(\gamma_0^{+}+\int_{\R^{+}}x\nu(dx)\right)t + \sigma \sqrt{\frac{2}{\pi}}\sqrt{t}.
\end{equation*}
\end{prop}

\begin{pf} \itshape proposition~\ref{levy_supremum_error}. \upshape
We will first prove the second result of the proposition. We have (see (\ref{levyactiviteinfinievf}))
\begin{eqnarray*}
\sup_{0\leq s\leq t}X_s&=&\sup_{0\leq s\leq t}\left(\gamma_0s+\sigma B_s + \int_{x\in\R,\tau\in[0,s]} xJ_X(dx \times d\tau)\right)
\\&\leq&\gamma_0^{+}t+\sigma \sup_{0\leq s\leq t}B_s + \int_{x\in\R^{+},\tau\in[0,t]} xJ_X(dx \times d\tau).
\end{eqnarray*}
So
\begin{eqnarray*}
\E\sup_{0\leq s\leq t}X_s&\leq&\gamma_0^{+}t+\sigma \E\sup_{0\leq s\leq t}B_s + t\int_{\R^{+}}x\nu(dx).
\end{eqnarray*}
By the reflexion theorem, we know that $\sup_{0 \leq s \leq t}B_s$ has the same distribution as $|B_t|$. Therefore
\begin{eqnarray*}
\E \sup_{0 \leq s \leq t}B_s =\E |B_t|=\sqrt{\frac{2}{\pi}}\sqrt{t}.
\end{eqnarray*}
Hence
\begin{eqnarray*}
\E\left(\sup_{0 \leq s \leq t} X_s \right) &\leq& \left(\gamma_0^{+}+\int_{\R^{+}}x\nu(dx)\right)t + \sigma \sqrt{\frac{2}{\pi}}\sqrt{t}.
\end{eqnarray*}
Consider now the general case. We define the process $\left(R_t\right)_{t \geq 0}$ by
\begin{eqnarray*} 
R_t=\lim_{\epsilon \downarrow 0} \widetilde{X}_t^{\epsilon}=\lim_{\epsilon \downarrow 0}\int_{\epsilon \leq |x| \leq 1,s\in[0,t]} x\widetilde{J}_X(dx \times ds).
\end{eqnarray*} 
We have, using (\ref{levy_Ito}),
\begin{eqnarray*} 
\E\left(\sup_{0 \leq s \leq t} X_s \right)&\leq& \E \sup_{0 \leq s \leq t} \left(\gamma s+\sigma B_s+X_s^l\right)+\E \sup_{0 \leq s \leq t}\left(R_s\right).
\end{eqnarray*}
The process $\left(\gamma s+\sigma B_s+X_s^l\right)_{t \geq 0}$ has finite activity and the support of 
its  L\'evy measure does not intersect $[-1,1]$, so
\begin{eqnarray*}
\E \sup_{0 \leq s \leq t} \left(\gamma s+\sigma B_s+X_s^l\right)\leq\left(\gamma^{+}+\int_{{x>1}}x\nu(dx)\right)t + \sigma \sqrt{\frac{2}{\pi}}\sqrt{t}.
\end{eqnarray*}
Besides, using the Cauchy-Schwarz and Doob inequalities (note that $R$ is a martingale) we get
\begin{eqnarray*}
\E \sup_{0 \leq s \leq t} \left(R_s\right) &\leq&2\sqrt{t\int_{|x|\leq1} x^2 \nu(dx)}.
\end{eqnarray*}
Hence
\begin{eqnarray*}
\E\left(\sup_{0 \leq s \leq t} X_s \right)\leq\left(\gamma^{+}+\int_{x>1}x\nu(dx)\right)t + \left(\sigma \sqrt{\frac{2}{\pi}}+2\sqrt{\int_{|x|\leq1} x^2 \nu(dx)}\right)\sqrt{t}.
\end{eqnarray*}
\end{pf}

\begin{pf} \itshape proposition~\ref{spitzer}. \upshape
By Proposition~\ref{levy_supremum_error} we have
\begin{eqnarray}\label{ineq1}
\exists c_1,c_2>0,\quad \forall t\geq0,\quad \E\sup_{0 \leq s \leq t}X_s \leq c_1 t + c_2 \sqrt{t}. 
\end{eqnarray}
Thus
\begin{eqnarray*}
\frac{\E X_s^{+}}{s} &\leq& \frac{\E \sup_{0 \leq \tau \leq s}X_{\tau}}{s}
\\ &\leq& c_1+\frac{c_2}{\sqrt{s}}.
\end{eqnarray*}
Since $s \rightarrow \frac{1}{\sqrt{s}}$ is integrable on $[0,t]$, so is $s\rightarrow\frac{\E X_s^{+}}{s}$. For $s\in(0,t]$, define
\begin{eqnarray*}
f(s)&=&\frac{\E X_s^{+}}{s}
\\f_n(s)&=&\sum_{k=1}^n\indicatrice_{\left(\frac{(k-1)t}{n},\frac{kt}{n}\right]}(s)f\left(\frac{kt}{n}\right),
\end{eqnarray*}
so that
\begin{eqnarray*}
\sum_{k=1}^{n} \frac{\E X_{k\frac{t}{n}}^{+}}{k}=\frac{t}{n}\sum_{k=1}^{n}f\left(\frac{kt}{n}\right)=\int_0^t f_n(s)ds.
\end{eqnarray*}
We can prove that $f$ is continuous on $(0,t]$. We deduce that $\lim_{n\rightarrow +\infty}f_n=f$ $a.e.$ We also have for any $s \in (0,t]$ 
\begin{eqnarray*}
|f_n(s)| &\leq& \sum_{k=1}^n\indicatrice_{\left(\frac{(k-1)t}{n},\frac{kt}{n}\right]}(s)\left|f\left(\frac{kt}{n}\right)\right|
\\ &\leq& \sum_{k=1}^n\indicatrice_{\left(\frac{(k-1)t}{n},\frac{kt}{n}\right]}(s) \left(c_1+\frac{c_2}{\sqrt{\frac{kt}{n}}}\right)
\\ &\leq& c_1+\frac{c_2}{\sqrt{s}}.
\end{eqnarray*}
So, by dominated convergence, we have $\lim_{n\rightarrow +\infty}\sum_{k=1}^{n} \frac{\E X_{k\frac{t}{n}}^{+}}{k}=\int_0^t \frac{\E X_s^{+}}{s}ds$. On the other hand
\begin{eqnarray*}
\max_{k=0, \dots,n}X_{k\frac{t}{n}}&=&\max\left(0,X_{\frac{t}{n}},X_{2\frac{t}{n}},\dots,X_t\right)
\\&=&\max\left(X_{\frac{t}{n}}^{+},X_{2\frac{t}{n}}^{+},\dots,X_t^{+}\right).
\end{eqnarray*}
Note that, for
$k\geq 1$, we have $X_{k\frac{t}{n}}=\sum_{j=1}^k \left(X_{j\frac{t}{n}}-X_{(j-1)\frac{t}{n}}\right)$
and the random variables $\left(X_{j\frac{t}{n}}-X_{(j-1)\frac{t}{n}}\right)_{j\geq 1}$ are i.i.d. So by Spitzer's identity, we have
\begin{eqnarray*}
\E \max_{k=0, \dots,n}X_{k\frac{t}{n}}=\sum_{k=1}^n \frac{1}{k} \E X_{k\frac{t}{n}}^{+}.
\end{eqnarray*}
The sequence $\left(\max_{k=0, \dots,n}X_{k\frac{t}{n}}\right)_{n\geq 0}$ is dominated by $\sup_{0 \leq s \leq t} X_s$, so by using the dominated convergence theorem, we get
\begin{eqnarray*}
\E \sup_{0 \leq s \leq t} X_s &=&\E \lim_{n \rightarrow +\infty}\max_{k=0, \dots,n}X_{k\frac{t}{n}}
\\&=&\lim_{n \rightarrow +\infty}\E \max_{k=0, \dots,n}X_{k\frac{t}{n}}
\\&=& \lim_{n \rightarrow +\infty} \sum_{k=1}^n \frac{1}{k} \E X_{k\frac{t}{n}}^{+}
\\ &=& \int_0^t \frac{\E X_s^{+}}{s} ds.
\end{eqnarray*}
\end{pf}

\subsection{Case of finite activity L\'evy processes}
The use of proposition~\ref{spitzer} in the finite activity case, leads to the following theorem.
\begin{thm}\label{weak_error}
Let $X$ be a finite activity L\'evy process satisfying $\int_{x>1}x\nu(dx)<\infty$, $t >0$ and $n \in \N$. 
\begin{enumerate}
  \item  If $\sigma>0$, we have, for $n\rightarrow +\infty$,
\small
\begin{eqnarray*}
\E\left(M_t -M_t^n\right)&=&
\frac{1}{2n}\left(\frac{\gamma_0 t}{2}+\lambda t \E Y_1^{+} -\sigma \sqrt{t} \E \phi\left(\frac{\gamma_0}{\sigma}\sqrt{t} +
 \frac{\sum_{i=1}^{N_t}Y_i}{\sigma \sqrt{t}}\right)\right)
\\&& - \frac{1}{2n}\E \left(\gamma_0 t+ \sum_{i=1}^{N_t}Y_i\right) \Phi\left(\frac{\gamma_0}{\sigma}\sqrt{t} + \frac{\sum_{i=1}^{N_t}Y_i}{\sigma \sqrt{t}}\right)
 \\ &&-\frac{\sigma \sqrt{t}\zeta \left(\frac{1}{2}\right)}{\sqrt{2 \pi n}}+ o\left(\frac{1}{n}\right).
\end{eqnarray*}
\normalsize
Here, $\zeta$ is the zeta Riemann function and $\phi$ and $\Phi$ are the probality density function and the cumulative distribution function of the standard normal distribution.
  \item If $\sigma=0$, then $s\rightarrow \frac{\E X_s^{+}}{s}$ is absolutely continuous on $[0,t]$ and we have
\begin{equation*}
\E\left(M_t -M_t^n\right)=\frac{1}{2n}\left(\gamma_0^{+}t+\lambda t\E Y_1^{+}-\E X_t^{+}\right) + o\left(\frac{1}{n}\right)
\end{equation*}
when $n \rightarrow +\infty$.
\end{enumerate}
\end{thm}
Recall that in the case of Brownian motion, Broadie Glasserman and Kou  prove in \cite{broadie-glasserman-kou99} (cf. lemma 3) a result similar to the first point of the above theorem. In the case $\sigma=0$, if $Y_1$ have a continuous density function or $\gamma_0=0$, the error $o\left(\frac{1}{n}\right)$ is in fact $O\left(\frac{1}{n^2}\right)$ (see \cite{dia}). To prove Theorem~\ref{weak_error}, we need the following more or less elementary lemmas.
\begin{lemme}\label{rmq1lemme}
 Let $f\in C^2[0,t]$. Then
     \begin{eqnarray*}
	   \int_0^t \frac{1}{\sqrt{x}}f(\sqrt{x})dx&=& \frac{t}{n} \sum_{k=1}^n \frac{1}{\sqrt{\frac{kt}{n}}}f\left(\sqrt{\frac{kt}{n}}\right)-\frac{\sqrt{t}\zeta\left(\frac{1}{2}\right)f(0)}{\sqrt{n}} \\&&-\frac{\sqrt{t}f(\sqrt{t})-tf'(0)}{2n}+ o\left(\frac{1}{n}\right). 
     \end{eqnarray*}
\end{lemme}
\begin{lemme}\label{Imapp}
Let $f$ be an absolutely continuous function  on $[0,t]$, then we have
\begin{eqnarray*}
\int_0^t f(s) ds- \frac{t}{n} \sum_{k=1}^n f\left(\frac{kt}{n}\right)
&=&\frac{t}{2n}\left(f(0)-f(t)\right)+o\left(\frac{1}{n}\right).
\end{eqnarray*}
\end{lemme}
The proof of the previous lemma is based on  the following result.
\begin{lemme}\label{Im}
Let $h\in L^1([0,t])$, we define the sequence $(I_m(h))_{m\geq1}$ by
\begin{equation*}
 I_m(h)= \sum_{k=1}^m\int_{(k-1)\frac{t}{m}}^{k\frac{t}{m}}h(u)\left(u-(k-1)\frac{t}{m}\right)du.
\end{equation*}
Then we have 
\begin{equation*}
 \lim_{m\rightarrow\ +\infty}mI_m(h)= \frac{t}{2}\int_{0}^{t}h(u)du.
\end{equation*}
\end{lemme}

\begin{pf} \itshape lemma~\ref{Im}. \upshape 
Consider first the case where $h\in C([0,t])$. By the variable substitutions $v=u-(k-1)\frac{t}{m}$, then $w=mv$ we get
\begin{eqnarray*}
I_m(h)&=& \sum_{k=1}^m\int_{0}^{\frac{t}{m}}h\left(v+(k-1)\frac{t}{m}\right)vdv \\
&=& \sum_{k=1}^m\int_{0}^{t}h\left(\frac{w}{m}+(k-1)\frac{t}{m}\right)\frac{w}{m}\frac{dw}{m} \\
&=& \frac{1}{m}\int_{0}^{t}\frac{1}{m}\sum_{k=1}^mh\left(\frac{w}{m}+(k-1)\frac{t}{m}\right)wdw.
\end{eqnarray*}
But $h$ is continuous and for $w\in [0,t]$ we have $\frac{w}{m}+(k-1)\frac{t}{m}\in \left[(k-1)\frac{t}{m},k\frac{t}{m}\right]$, so
\begin{eqnarray*}
\lim_{m\rightarrow\ +\infty}\frac{t}{m}\sum_{k=1}^mh\left(\frac{w}{m}+(k-1)\frac{t}{m}\right)= \int_0^th(s)ds.
\end{eqnarray*}
Hence
\begin{eqnarray*}
\lim_{m\rightarrow\ +\infty}mI_m(h) &=& \int_{0}^{t}\left(\frac{1}{t}\int_0^th(s)ds\right)wdw\\
&=& \frac{t}{2}\int_0^th(s)ds.
\end{eqnarray*}
Consider now the case where $h$ is integrable on $[0,t]$. Then there exists a sequence of functions $(h_n)_{n\geq0}$ in $C([0,t])$ such that
\begin{equation*}
\lim_{n\rightarrow\ +\infty}\int_0^t|h(u)-h_n(u)|du=0.
\end{equation*}
So we have
\begin{eqnarray*}
u_m^n:&=&\left|mI_m(h_n)-m\sum_{k=1}^m\int_{(k-1)\frac{t}{m}}^{k\frac{t}{m}}h(u)\left(u-(k-1)\frac{t}{m}\right)du\right| 
\\&=&\left|m\sum_{k=1}^m\int_{(k-1)\frac{t}{m}}^{k\frac{t}{m}}(h_n(u)-h(u))\left(u-(k-1)\frac{t}{m}\right)du\right|\\
&\leq&m\sum_{k=1}^m\int_{(k-1)\frac{t}{m}}^{k\frac{t}{m}}|h_n(u)-h(u)|\left|u-(k-1)\frac{t}{m}\right|du\\
&\leq&t\sum_{k=1}^m\int_{(k-1)\frac{t}{m}}^{k\frac{t}{m}}|h_n(u)-h(u)|du\\
&\leq&t\int_{0}^{t}|h_n(u)-h(u)|du.
\end{eqnarray*}
The convergence (with respect to $m$) of $mI_m(h_n)$ is uniform. Hence by the limits inversion theorem
\begin{eqnarray*}
&&\lim_{m\rightarrow\ +\infty}\lim_{n\rightarrow\ +\infty}mI_m(h_n)=\lim_{n\rightarrow\ +\infty}\lim_{m\rightarrow\ +\infty}mI_m(h_n)\\
&&\Rightarrow \lim_{m\rightarrow\ +\infty}mI_m(h)=\lim_{n\rightarrow\ +\infty}\frac{t}{2}\int_0^th_n(u)du\\
&&\Rightarrow \lim_{m\rightarrow\ +\infty}mI_m(h)=\frac{t}{2}\int_0^th(u)du.
\end{eqnarray*}
\end{pf}

\begin{pf} \itshape lemma~\ref{Imapp}. \upshape
Let $h$ be the a.e. derivative of $f$. We have
\begin{eqnarray*}
\int_0^t f(s) ds- \frac{t}{n} \sum_{k=1}^n f\left(\frac{kt}{n}\right)
&=&\sum_{k=1}^n\int_{(k-1)\frac{t}{n}}^{k\frac{t}{n}} \left(f(s)-f\left(\frac{kt}{n}\right)\right)ds\\
&=&-\sum_{k=1}^n\int_{(k-1)\frac{t}{n}}^{k\frac{t}{n}}\int_s^{k\frac{t}{n}}h(u)duds\\
&=&-\sum_{k=1}^n\int_{(k-1)\frac{t}{n}}^{k\frac{t}{n}}\int_{(k-1)\frac{t}{n}}^uh(u)dsdu, \ \mbox{by Fubini.}
\end{eqnarray*}
Thus
\begin{eqnarray*}
\int_0^t f(s) ds- \frac{t}{n} \sum_{k=1}^n f\left(\frac{kt}{n}\right)
&=&-\sum_{k=1}^n\int_{(k-1)\frac{t}{n}}^{k\frac{t}{n}}h(u)\left(u-(k-1)\frac{t}{n}\right)du\\
&=&-\frac{t}{2n}\int_0^th(u)du+o\left(\frac{1}{n}\right), \ \mbox{by lemma~\ref{Im}}\\
&=&-\frac{t}{2n}\left(f(t)-f(0)\right)+o\left(\frac{1}{n}\right)\\
&=&\frac{t}{2n}\left(f(0)-f(t)\right)+o\left(\frac{1}{n}\right).
\end{eqnarray*}
\end{pf}

\begin{pf} \itshape lemma~\ref{rmq1lemme}. \upshape
We consider first the case $t=1$. The case $t\neq 1$ will be deduced by a variable substitution. We have
\begin{equation*}
\frac{1}{\sqrt{x}}f(\sqrt{x})=\frac{f(0)}{\sqrt{x}}+\frac{f\left(\sqrt{x}\right)-f(0)}{\sqrt{x}}.
\end{equation*}
Set
\begin{equation*}
g(x)=\frac{f\left(\sqrt{x}\right)-f(0)}{\sqrt{x}}.
\end{equation*}
The function $g$ can be extended to a continuous function on $[0,1]$, and \\$\lim_{x\rightarrow0}g(x)=f'(0)$. Furthermore $g$ is differentiable on $(0,1]$ and
\begin{equation*}
g'(x)=\frac{f(0)-f\left(\sqrt{x}\right)+\sqrt{x}f'\left(\sqrt{x}\right)}{2x^{\frac{3}{2}}}.
\end{equation*}
The function $g'$ is integrable on $[0,1]$, so $g$ is absolutely continuous. Thus 
\begin{eqnarray*}
\epsilon_n(f)
&=&\int_0^1\frac{f(0)}{\sqrt{x}}dx+\int_0^1 g(x)dx- \frac{1}{n} \sum_{k=1}^n\frac{f(0)}{\sqrt{\frac{k}{n}}}-\frac{1}{n} \sum_{k=1}^n g\left(\frac{k}{n}\right)\\
&=&f(0)\left(\int_0^1\frac{1}{\sqrt{x}}dx- \frac{1}{n} \sum_{k=1}^n\frac{1}{\sqrt{\frac{k}{n}}}\right)
+\left(\int_0^1 g(x)dx-\frac{1}{n} \sum_{k=1}^n g\left(\frac{k}{n}\right)\right).
\end{eqnarray*}
By using \cite{knopp} (see p.538) and lemma~\ref{Imapp}, we get
\begin{eqnarray*}
\epsilon_n(f)
&=&f(0)\left(-\frac{\zeta \left(\frac{1}{2}\right)}{\sqrt{n}} -\frac{1}{2n}+O\left(\frac{1}{n^2}\right)\right)+\frac{g(0)}{2n}-\frac{g(1)}{2n}+ o\left(\frac{1}{n}\right) \\
&=&-\frac{\zeta \left(\frac{1}{2}\right)}{\sqrt{n}}f(0)-\frac{f(0)}{2n}-\frac{f(1)-f'(0)-f(0)}{2n}+ o\left(\frac{1}{n}\right) \\
&=&-\frac{\zeta \left(\frac{1}{2}\right)f(0)}{\sqrt{n}} -\frac{f(1)-f'(0)}{2n}+ o\left(\frac{1}{n}\right).   
\end{eqnarray*}
\end{pf}

\begin{pf} \itshape theorem~\ref{weak_error}. \upshape
We know by theorem~\ref{spitzer} that
\begin{eqnarray*}
\E\left(\sup_{0 \leq s \leq t} X_s -\max_{k=0, \dots,n}X_{k\frac{t}{n}}\right)=\int_0^t \frac{\E X_s^{+}}{s} ds- \frac{t}{n} \sum_{k=1}^n \frac{\E X_{k\frac{t}{n}}^{+}}{\frac{kt}{n}}.
\end{eqnarray*}
So we need to study the smoothness of the function $s\mapsto {\E X_s^{+}/s}$ and conclude with  lemmas \ref{rmq1lemme} and \ref{Imapp}. 

{\em Case $1$} : $\sigma>0$ and $\E Y_1^{+}<\infty$.
\\Let $U$ be a normal r.v. with mean $\gamma$ and variance $\sigma^2$. 
By an easy computation we get
\begin{eqnarray*}
&&\E U^{+}=\sigma \phi\left(\frac{\gamma}{\sigma}\right) + \gamma\Phi\left(\frac{\gamma}{\sigma}\right).
\end{eqnarray*}
So, for any $s>0$, we have, by conditionning with respect to the jump part of the process $X$, \small
\begin{eqnarray*}
\E \frac{X_s^{+}}{s}=\E \frac{\sigma}{\sqrt{s}} \phi\left(\frac{\gamma_0}{\sigma}\sqrt{s}+ \frac{\sum_{i=1}^{N_s}Y_i}{\sigma \sqrt{s}}\right) + \E\left(\gamma_0+\frac{\sum_{i=1}^{N_s}Y_i}{s}\right)\Phi\left(\frac{\gamma_0}{\sigma}\sqrt{s}+ \frac{\sum_{i=1}^{N_s}Y_i}{\sigma \sqrt{s}}\right).
\end{eqnarray*}\normalsize
Let $f$ and $g$ be the functions defined by
\begin{eqnarray*}
&&f(s)=\E\phi\left(\frac{\gamma_0}{\sigma}s+ \frac{\sum_{i=1}^{N_{s^2}}Y_i}{\sigma s}\right)
\\&&g(s)=\E\left(\frac{\gamma_0}{\sigma} s+\frac{\sum_{i=1}^{N_{s^2}}Y_i}{\sigma s}\right)\Phi\left(\frac{\gamma_0}{\sigma}s+ \frac{\sum_{i=1}^{N_{s^2}}Y_i}{\sigma s}\right),
\end{eqnarray*}
so that
\begin{eqnarray*}
\E \frac{X_s^{+}}{s}=\frac{\sigma}{\sqrt{s}}f\left(\sqrt{s}\right) + \frac{\sigma}{\sqrt{s}}g\left(\sqrt{s}\right).
\end{eqnarray*}
If $f$ and $g$ can be extended as $C^2$ functions on $[0,t]$ then, using lemma~\ref{rmq1lemme}, we get the first part of the theorem. By \cite{ContTankov},  proposition 9.5, we have
\begin{eqnarray*}
f(s)&=&\E s^{2N_1}e^{-\lambda(s^2-1)}\phi\left(\frac{\gamma_0}{\sigma}s+ \frac{\sum_{i=1}^{N_1}Y_i}{\sigma s}\right).
\end{eqnarray*}
So, the function $f$ has the same regularity as $\tilde{f}$ defined by
\begin{eqnarray*}
\tilde{f}(s)=\E s^{2N_1}\phi\left(\mu s+ \frac{\sum_{i=1}^{N_1}Y_i}{\sigma s}\right),
\end{eqnarray*}
where $\mu=\frac{\gamma_0}{\sigma}$.  For $x\in \R$, we define the function
\begin{eqnarray*}
s\mapsto h(s,x)=\phi\left(\mu s+ \frac{x}{s}\right).
\end{eqnarray*}
We then have
\begin{eqnarray*}
\tilde{f}(s)=\E s^{2N_1}h\left(s,\frac{\sum_{i=1}^{N_1}Y_i}{\sigma}\right).	
\end{eqnarray*}
Note that
\begin{eqnarray*}
0\leq h\left(s,x\right)\leq\frac{1}{\sqrt{2\pi}},
\end{eqnarray*}
and
\begin{eqnarray*}
h\left(s,x\right)&=&\frac{1}{\sqrt{2\pi}}\exp\left(-\frac{1}{2}\left(\mu s+\frac{x}{s}\right)^2\right)
\\&=&\frac{1}{\sqrt{2\pi}}\exp\left(-\frac{\mu^2s^2}{2}\right)\exp\left(-\mu x-\frac{x^2}{2s^2}\right).
\end{eqnarray*}
Using the inequality $-\mu x\leq\mu^2 s^2+\frac{x^2}{4s^2}$, we get
\begin{equation}\label{maj_h}
h\left(s,x\right)\leq\frac{1}{\sqrt{2\pi}}\left(e^{\frac{\mu^2s^2}{2}}e^{-\frac{x^2}{4s^2}}\wedge 1\right).
\end{equation}
Moreover, we have
\begin{eqnarray*}
\frac{\partial}{\partial s} h\left(s,x\right)&=&\left(\frac{x^2}{s^3}-\mu^2s\right)h(s,x)
\end{eqnarray*}
and
\begin{eqnarray*}
\frac{\partial^2}{\partial s^2} h\left(s,x\right)&=&\left(-\frac{3x^2}{s^4}-\mu^2\right)\phi\left(\mu s+ \frac{x}{s}\right)+\left(\frac{x^2}{s^3}-\mu^2s\right)^2\phi\left(\mu\sqrt{s}+ \frac{x}{\sqrt{s}}\right).
\end{eqnarray*}
Using \eqref{maj_h}, we get
\begin{eqnarray*}
\left|\frac{\partial}{\partial s} h\left(s,x\right)\right|&\leq&\frac{\mu^2s}{2\sqrt{2\pi}}+\frac{x^2}{s^3\sqrt{2\pi}}e^{\frac{\mu^2s^2}{2}}e^{-\frac{x^2}{4s^2}}
\\&\leq&\frac{\mu^2s}{2\sqrt{2\pi}}+\frac{C_1\indicatrice_{\left\{x\neq0\right\}}}{s}e^{\frac{\mu^2s}{2}},
\end{eqnarray*}
where $C_1=\sup_{y>0}\left(\frac{y^2e^{-\frac{y^2}{4}}}{\sqrt{2\pi}}\right)$. Using \eqref{maj_h} again  
and the fact that \\$\left(\frac{x^2}{s^3}-\mu^2s\right)^2\leq2\left(\frac{x^4}{s^6}+\mu^4s^2\right)$, we 
obtain
\begin{eqnarray*}
\left|\frac{\partial^2}{\partial s^2} h\left(s,x\right)\right|&\leq&\left(\mu^2 +2\mu^4s^2\right)h(s,x)+
\left(\frac{3x^2}{s^4}+2\frac{x^4}{s^6}\right)h(s,x)
\\&\leq&\frac{\mu^2 +2\mu^4s^2}{\sqrt{2\pi}}+\frac{C_2\indicatrice_{x\neq0}}{s^2}e^{\frac{\mu^2s^2}{2}},
\end{eqnarray*}
where $C_2=\sup_{y>0}\left(\frac{3y^2+2y^4}{\sqrt{2\pi}}e^{-\frac{y^2}{4}}\right)$. Hence \small
\begin{eqnarray*}
\frac{\partial}{\partial s}\left(s^{2N_1} h\left(s,\frac{\sum_{i=1}^{N_1}Y_i}{\sigma}\right)
\right)&=&2N_1s^{2N_1-1}h\left(s,\frac{\sum_{i=1}^{N_1}Y_i}{\sigma}\right)\\&&+s^{2N_1}\frac{\partial}{\partial 
s}h\left(s,\frac{\sum_{i=1}^{N_1}Y_i}{\sigma}\right).
\end{eqnarray*}\normalsize
Thus \small
\begin{eqnarray*}
\left|\frac{\partial}{\partial s}\left(s^{2N_1} h\left(s,\frac{\sum_{i=1}^{N_1}Y_i}{\sigma}\right)\right)\right|&\leq&\frac{2N_1s^{2N_1-1}}{\sqrt{2\pi}}+\frac{\mu^2s^{2N_1-1}}{\sqrt{2\pi}}+C_1\indicatrice_{\left\{N_1>0\right\}}s^{2N_1-1}e^{\frac{\mu^2s^2}{2}}.
\end{eqnarray*}\normalsize
We deduce that $\tilde{f}$ is continuously differentiable, and
\begin{eqnarray*}
\tilde{f}'(s)=\E\left[2N_1s^{2N_1-1}h\left(s,\frac{\sum_{i=1}^{N_1}Y_i}{\sigma}\right)+s^{2N_1}\frac{\partial}{\partial s}h\left(s,\frac{\sum_{i=1}^{N_1}Y_i}{\sigma}\right)\right].
\end{eqnarray*}
Similarly,
\begin{eqnarray*}
&&\frac{\partial^2}{\partial s^2}\left(s^{2N_1} h\left(s,\frac{\sum_{i=1}^{N_1}Y_i}{\sigma}\right)\right)= 2N_1\left(2N_1-1\right)s^{2N_1-2}h\left(s,\frac{\sum_{i=1}^{N_1}Y_i}{\sigma}\right)
\\&&\quad \quad \quad \quad \quad \quad +4N_1s^{2N_1-1}\frac{\partial}{\partial s}h\left(s,\frac{\sum_{i=1}^{N_1}Y_i}{\sigma}\right)
+s^{2N_1}\frac{\partial^2}{\partial s^2}h\left(s,\frac{\sum_{i=1}^{N_1}Y_i}{\sigma}\right).
\end{eqnarray*}
But
\begin{eqnarray*}
&&\left|2N_1\left(2N_1-1\right)s^{2N_1-2}h\left(s,\frac{\sum_{i=1}^{N_1}Y_i}{\sigma}\right)\right|\leq
\frac{2N_1\left(2N_1-1\right)s^{2N_1-2}}{\sqrt{2\pi}}\\
&&\left|42N_1s^{2N_1-1}\frac{\partial}{\partial s}\left(s,\frac{\sum_{i=1}^{N_1}Y_i}{\sigma}\right)\right|\leq
\frac{4N_1\left(2N_1-1\right)s^{N_1}}{\sqrt{2\pi}}\\&&\quad \quad \quad \quad \quad \quad \quad \quad \quad \quad \quad \quad \quad \quad \quad  +4N_1\left(2N_1-1\right)s^{2N_1-2}C_1
\indicatrice_{\left\{N_1>0\right\}}e^{\frac{\mu^2s^2}{2}}\\
&&\left|s^{2N_1}\frac{\partial^2}{\partial s^2}h\left(s,\frac{\sum_{i=1}^{N_1}Y_i}{\sigma}\right)\right|\leq
\frac{\mu^2 +2\mu^4s^2}{\sqrt{2\pi}}s^{2N_1}+C_2\indicatrice_{N_1>0}s^{2N_1-2}e^{\frac{\mu^2s^2}{2}}.
\end{eqnarray*}
We deduce that $\tilde{f}$ is twice differentiable on $[0,t]$ and \small
\begin{eqnarray*}
\tilde{f}{''}(s)&=&\E\left[2N_1\left(2N_1-1\right)s^{2N_1-2}h\left(s,\frac{\sum_{i=1}^{N_1}Y_i}{\sigma}\right)+4N_1s^{2N_1-1}\frac{\partial}{\partial s}h\left(s,\frac{\sum_{i=1}^{N_1}Y_i}{\sigma}\right)\right]
\\&&+\E\left[s^{2N_1}\frac{\partial^2}{\partial s^2}h\left(s,\frac{\sum_{i=1}^{N_1}Y_i}{\sigma}\right)\right].
\end{eqnarray*}\normalsize
Hence $f$ is in $C^{2}[0,t]$ and we verify that $f(0)=\frac{1}{\sqrt{2\pi}}$ and $f'(0)=0$. On the other 
hand the function $g$ can be written in the following form (see \cite{ContTankov}, proposition 9.5)
\begin{eqnarray*}
g(s)=\E s^{2N_1}e^{-\lambda(s^2-1)}\left(\frac{\gamma_0}{\sigma}s+\frac{\sum_{i=1}^{N_{1}}Y_i}{\sigma s}\right)\Phi\left(\frac{\gamma_0}{\sigma}s+ \frac{\sum_{i=1}^{N_{1}}Y_i}{\sigma s}\right).
\end{eqnarray*}
With the same reasoning we could prove that $g$ is in $C^2[0,t]$, and satisfies $g(0)=0$ and $g{'}(0)=\frac{\lambda\E Y_1^{+}}{\sigma}+\frac{\gamma_0}{2\sigma}$. This proves the first part of the theorem.

{\em  Case $2$: } $\sigma=0$ and $\E Y_1^{+}<\infty$. \\
We have
\begin{eqnarray*}
\frac{\E X_s^{+}}{s}=\gamma_0^{+}e^{-\lambda s}+ e^{-\lambda s}\sum_{n=1}^{+\infty} \frac{\lambda^n s^{n-1}}{n!} \E \left(\gamma_0s+\sum_{i=1}^{n}Y_i \right)^{+}.
\end{eqnarray*}
Observe that, for any positive integer $n$,
 the function  $s\mapsto \E \left(\gamma_0 s + \sum_{i=1}^{n}Y_i\right)^{+}$ is absolutely continuous. So 
 is
$s\mapsto \frac{\lambda^n s^{n-1}}{n!}\E \left(\gamma_0 s + \sum_{i=1}^{n}Y_i\right)^{+}$. If we call $h_n$ its a.e. derivative,  then, for any $n\geq2$,
\[
h_n(s)=\gamma_0\frac{\lambda^n s^{n-1}}{n!}\P\left(\gamma_0 s+\sum_{i=1}^{n}Y_i\geq 0\right)+
\frac{n-1}{n!}\lambda^n s^{n-2}\E \left(\gamma_0 s + \sum_{i=1}^{n}Y_i\right)^{+},
\]
so that, for $s\in [0,t]$,
\[
 |h_n(s)|\leq |\gamma_0|\frac{\lambda^n t^{n-1}}{n!}+\frac{n-1}{n!}\lambda^n t^{n-2}\left(|\gamma_0|t+n\E Y_1^{+}\right).
\]
Hence the normal convergence of $\sum h_n$ on $[0,t]$, and thus the absolute continuity of $\frac{\E X_s^{+}}{s}$ on $[0,t]$. So, by proposition~\ref{spitzer} and lemma~\ref{Imapp},
\begin{eqnarray*}
\E\left(\sup_{0 \leq s \leq t} X_s -\max_{k=0, \dots,n}X_{k\frac{t}{n}}\right)&=&\int_0^t \frac{\E X_s^{+}}{s} ds- \frac{t}{n} \sum_{k=1}^n \frac{\E X_{k\frac{t}{n}}^{+}}{\frac{kt}{n}}\\
&=&\frac{t}{2n}\left(\lim_{s\rightarrow 0^{+}}\frac{\E X_s^{+}}{s}-\frac{\E X_t^{+}}{t}\right)+o\left(\frac{1}{n}\right)\\
&=&\frac{1}{2n}\left(\left(\gamma_0^{+}+\lambda\E Y_1^{+}\right)t-E X_t^{+}\right)+o\left(\frac{1}{n}\right)\\
&=&\frac{1}{2n}\left(\gamma_0^{+}t+\lambda t\E Y_1^{+}-E X_t^{+}\right)+o\left(\frac{1}{n}\right).
\end{eqnarray*}
\end{pf}

\subsection{Case of infinite activity L\'evy processes}
In the case of L\'evy processes with infinite activity, we cannot use \eqref{levyactivitefinie}. So the method used in theorem~\ref{weak_error} does not work anymore and we must use another approach.
\begin{thm}\label{error}
Let $X$ be an integrable L\'evy process with generating triplet $(\gamma,\sigma^2,\nu)$. Then
\begin{enumerate}
   \item If $\sigma>0$
\begin{eqnarray*}
\E\left(M_t -M_t^n\right)=O\left(\frac{1}{\sqrt{n}}\right).
\end{eqnarray*}
   \item If $\sigma=0$
\begin{eqnarray*}
\E\left(M_t -M_t^n\right)=o\left(\frac{1}{\sqrt{n}}\right).
\end{eqnarray*}
   \item If $\sigma=0$ and $\int_{|x|\leq1}|x|\nu(dx)<\infty$
\begin{eqnarray*}
\E\left(M_t -M_t^n\right)=O\left(\frac{\log(n)}{n}\right).
\end{eqnarray*}
\end{enumerate}
\end{thm}

To prove the result $2$ of theorem~\ref{error}, we will use the lemma below.
\begin{lemme}\label{errorlemme2}
Let $X$ be an integrable L\'evy process with generating triplet $(\gamma,0,\nu)$. Then we have
\begin{eqnarray*}
\E X_t^{+}=o\left(\sqrt{t}\right)
\end{eqnarray*}
when $t\rightarrow 0$.
\end{lemme}

The proof of this lemma is quite standard, and is left to the reader. For more details, see \cite{dia}.

\begin{pf} \itshape theorem~\ref{error}.  \upshape
With the notation $\delta=\frac{t}{n}$, we have, using  proposition~\ref{spitzer},
\begin{eqnarray*}
\E\left(M_t -M_t^n\right)&=&\int_0^t \frac{\E X_s^{+}}{s} ds- \sum_{k=1}^n 
\frac{\E X_{k\delta}^{+}}{k}\\
&=&\sum_{k=1}^{n}\int_{(k-1)\delta}^{k\delta} \frac{\E X_s^{+}}{s}- 
            \sum_{k=1}^{n}\int_{(k-1)\delta}^{k\delta}\frac{\E X_{k\delta}^{+}}{k\delta}ds\\
&=&\int_{0}^{\delta} \left(\frac{\E X_s^{+}}{s}- \frac{\E X_{\delta}^{+}}{\delta}\right)ds +
      \sum_{k=2}^{n}\int_{(k-1)\delta}^{k\delta} \left(\frac{\E X_s^{+}}{s}- 
                          \frac{\E X_{k\delta}^{+}}{k\delta}\right)ds.
\end{eqnarray*}
We call $u(\delta)$ (respectively $v(\delta)$) the first (respectively the second) 
term on the right of the last equality. We easily deduce from Proposition~\ref{levy_supremum_error}
that, if $\sigma>0$, $u(\delta)=O(\sqrt{\delta})$ and, if $\sigma=0$ and 
$\int_{|x|\leq1}|x|\nu(dx)<\infty$, $u(\delta)=O({\delta})$. We also have
\begin{eqnarray*}
\frac{u(\delta)}{\sqrt{\delta}}&=&\int_0^\delta \frac{\E X_s^{+}}{s\sqrt{\delta}}ds -
\frac{\E X_\delta^{+}}{\sqrt{\delta}}\\
&=&\int_{0}^{1} \frac{1}{\sqrt{s}}\frac{\E X_{s\delta}^{+}}{\sqrt{s\delta}}
-\frac{\E X_\delta^{+}}{\sqrt{\delta}},
\end{eqnarray*}
and we easily deduce from Lemma~ \ref{errorlemme2} that, if $\sigma=0$,
$u(\delta)=o(\sqrt{\delta})$. 

We now study $v(\delta)$. For $s\geq 0$, let  $\tilde{X}_s=X_s-\alpha s$, where $\alpha=\E X_1$. 
Then, $\tilde{X}$ is a martingale and, for a fixed $s\geq0$, 
$\left(\tilde{X}_{\tau}+\alpha s\right)^{+}_{\tau\geq0}$ is a submartingale, 
because $x\rightarrow x^{+}$ is a convex function. So, for $s\in[(k-1)\delta,\delta]$,
\[
 \E X_s^{+}=\E\left(\tilde{X}_{s}+\alpha s\right)^{+}\leq\E\left(\tilde{X}_{k\delta}+\alpha s\right)^{+}\!.
\]
Hence
\begin{eqnarray*}
v(\delta)&=&\sum_{k=2}^{n}\int_{(k-1)\delta}^{k\delta} \left(\frac{\E X_s^{+}}{s}-
 \frac{\E X_{k\delta}^{+}}{k\delta}\right)ds\\
 &\leq&\sum_{k=2}^{n}\int_{(k-1)\delta}^{k\delta} \left(\frac{\E \left(\tilde{X}_{k\delta}+\alpha s\right)^{+}}{s}- \frac{\E \left(\tilde{X}_{k\delta}+\alpha k\delta\right)^{+}}{k\delta}\right)ds
\\&=&\sum_{k=2}^{n}\int_{(k-1)\delta}^{k\delta} \E \left(\tilde{X}_{k\delta}+\alpha k\delta\right)^{+}\left(\frac{1}{s}- \frac{1}{k\delta}\right)ds
\\&&+
\sum_{k=2}^{n}\int_{(k-1)\delta}^{k\delta} \frac{\E \left(\tilde{X}_{k\delta}+\alpha s\right)^{+}-E \left(\tilde{X}_{k\delta}+\alpha k\delta\right)^{+}}{s}ds.
\end{eqnarray*}
Using the inequality  $|x^{+}-y^{+}|\leq |x-y|$, we get
\begin{eqnarray*}
v(\delta)&\leq&\sum_{k=2}^{n} \E X_{k\delta}^{+}\left(\log\left(\frac{k}{k-1}\right)- \frac{1}{k}\right)+
\sum_{k=2}^{n}\int_{(k-1)\delta}^{k\delta} \frac{|\alpha|\left(k\delta-s\right)}{s}ds\\
&=&\sum_{k=2}^{n} \E X_{k\delta}^{+}\left(\log\left(1+\frac{1}{k-1}\right)- \frac{1}{k}\right)ds+
\sum_{k=2}^{n}\int_{(k-1)\delta}^{k\delta} |\alpha|\left(\frac{k\delta}{s}-1\right)ds
\\&\leq&\sum_{k=2}^{n} \E X_{k\delta}^{+}\left(\frac{1}{k-1}- \frac{1}{k}\right)+
\sum_{k=2}^{n} |\alpha|\delta\left(k\log\left(\frac{k}{k-1}\right)-1\right)\\
    &\leq&\sum_{k=2}^{n} \E X_{k\delta}^{+}\frac{1}{k(k-1)}+
 |\alpha|\delta\sum_{k=2}^{n}\left(\frac{k}{k-1}-1\right)\\
 &=&\sum_{k=2}^{n}\E X_{k\delta}^{+}\frac{1}{k(k-1)}+
 |\alpha|\delta\sum_{k=2}^{n}\frac{1}{k-1}.
\end{eqnarray*}
Now, if $\sigma=0$ and $\int_{|x|\leq1}|x|\nu(dx)<\infty$, we know from 
Proposition~\ref{levy_supremum_error} that $\E X_{k\delta}^{+}\leq Ck\delta$ for some $C>0$, so that
\begin{eqnarray*}
v(\delta)&\leq&C\delta\sum_{k=2}^{n}\frac{1}{k-1}+|\alpha|\delta\sum_{k=2}^{n}\frac{1}{k-1}
\\&\leq&\left(C\delta+|\alpha|\right)(1+\log(n-1))
\\&=&O\left(\frac{\log(n)}{n}\right),
\end{eqnarray*}
so that the last statement of the Theorem is proved.

For the other cases, let $f(s)=\E(X_s^+)/\sqrt{s}$, so that
\[
\sum_{k=2}^{n}\E X_{k\delta}^{+}\frac{1}{k(k-1)}=\sqrt{\delta}\sum_{k=2}^{n}
                 f(k\delta)\frac{1}{\sqrt{k}(k-1)}.
\]
We know from Proposition~\ref{levy_supremum_error} that $f$ is bounded on $[0,t]$, so that
the first statement of Theorem~\ref{error} 
now follows from the convergence of the series $\sum 1/k^{3/2}$.

In order to prove the second statement (i.e. the case $\sigma=0$), we observe that
$\sum_{k=2}^n f(k\delta)\frac{1}{\sqrt{k}(k-1)}$ goes to $0$ as $n\to \infty$, as follows easily
from $\lim_{s\to 0}f(s)=0$ (cf. Lemma~\ref{errorlemme2}). 
\end{pf}

\begin{rmq}\label{errorrmq} \rm
The second result of theorem~\ref{error} is optimal in the following sense: for any $\epsilon>0$, there exists a L\'evy process $X$ satisfying $\sigma=0$, such that
\begin{eqnarray*}
 \lim_{n\rightarrow +\infty}n^{\frac{1}{2}+\epsilon}\E\left(M_t -M_t^n\right)=+\infty.
\end{eqnarray*}
More precisely, if $X$ is a stable process of order $\alpha$, with $\alpha\in(1,2)$, we have
\[
\lim_{n\to\infty}n^{1/\alpha}\E\left(M_t -M_t^n\right)=
    -t^{1/\alpha}\zeta\left(1-\frac{1}{\alpha}\right)\E X_1^+.
\] 
The proof can be found in \cite{dia}.
\end{rmq}

In the finite variation case, with a stronger assumption, we extend the results on compound Poisson processes which we get in the previous section, to infinite activity case.
\begin{thm}\label{error2}
Let $X$ be an integrable L\'evy process with generating triplet $(\gamma,0,\nu)$. Suppose that
\\$\int_{|x|\leq1}|x|\left|\log\left(|x|\right)\right|\nu(dx)<\infty$ and $\nu(\R)=+\infty$, then
\begin{eqnarray*}
\E\left(M_t -M_t^n\right)=\left(\left(\gamma_0^{+} +\int_{\R}x^+\nu(dx)\right)t-\E X_t^{+}\right)\frac{1}{2n}
+o\left(\frac{1}{n}\right)\!.
\end{eqnarray*}
\end{thm}

\begin{lemme}\label{error2lemme2}
If $X$ is a finite variation L\'evy process with infinite activity  and $\gamma_0\neq0$, then
\begin{equation}\label{error2eq}
\int_0^tds\int_0^1du\frac{1}{s}\left|\P\left[X_s\geq0\right]-\P\left[X_{su}\geq0\right]\right|<\infty.
\end{equation}
\end{lemme}

\begin{pf} \itshape  lemma~\ref{error2lemme2}.  \upshape
We first consider  the case $\gamma_0<0$.
Recall that, since $X$ has finite variation, we have, with probability one, $\lim_{t\to 0}\frac{X_t}{t}=\gamma_0$, therefore $\P(R_0>0)=1$, where 
\[
R_0=\inf\{t>0\;|\; X_t>0\},
\]
 and
$\int_0^ts^{-1}{\P(X_s>0)}ds<\infty$
(see \cite{sato}, Section 47, especially Theorem 47.2). Set
\begin{eqnarray*}
I=\int_0^tds\int_0^1du\frac{1}{s}\left|\P\left[X_s\geq0\right]-\P\left[X_{su}\geq0\right]\right|
\end{eqnarray*}
Note that, since $X$ has infinite activity, we have $\P(X_s=0)=0$, for all $s>0$
(see \cite{sato}, Theorem 27.4), so that
\begin{eqnarray*}
I
&\leq&\int_0^t\frac{1}{s}\P\left[X_s\geq0\right]ds+\int_0^tds\int_0^1du\frac{1}{s}\P\left[X_{su}\geq0\right]\\
&=&\int_0^t\frac{1}{s}\P\left(X_s>0\right)ds+\int_0^tds\int_0^1du\frac{1}{s}\P\left(X_{su} > 0\right).
\end{eqnarray*}
So, we need to prove that $\int_0^tds\int_0^1dus^{-1}\P\left(X_{su} > 0\right)<\infty$.
We have
\begin{eqnarray*}
\int_0^tds\int_0^1du\frac{1}{s}\P\left[X_{su}>0\right]&=&\int_0^tds\int_0^sdu\frac{1}{s^2}
        \P\left[X_{u}>0\right]\\
        &=&\int_0^t\frac{1}{s^2}\left(\int_0^s\P\left[X_{u}>0\right]du\right)ds\\
        &=&\left[-\frac{1}{s}\left(\int_0^s\P\left[X_{u}>0\right]du\right)\right]_0^t+\int_0^t\frac{1}{s}\P\left[X_{s}>0\right]ds.
\end{eqnarray*}
But, for any $s>0$,
\begin{eqnarray*}
\left|\frac{1}{s}\left(\int_0^s\P\left[X_{u}\geq0\right]du\right)\right|\leq1
\end{eqnarray*}
So,  using again $\int_0^ts^{-1}{\P(X_s>0)}ds<\infty$, we conclude that
\begin{eqnarray*}
\int_0^tds\int_0^1du\frac{1}{s}\P\left[X_{su}\geq0\right]<\infty.
\end{eqnarray*}
Consider now $\gamma_0>0$. Let $\tilde{X}$ be the  dual process of $X$ (e.g. $\tilde{X}=-X$). 
Then $\gamma_0^{\tilde{X}}=-\gamma_0$, and so $\gamma_0^{\tilde{X}}<0$. Thus
\begin{eqnarray*}
I
&=&\int_0^tds\int_0^1du\frac{1}{s}\left|\P\left[X_s<0\right]-\P\left[X_{su}<0\right]\right|
\\&=&\int_0^tds\int_0^1du\frac{1}{s}\left|\P\left[\tilde{X}_s\geq0\right]-\P\left[\tilde{X}_{su}\geq0\right]\right|
\\&<&\infty.
\end{eqnarray*}
\end{pf}

\begin{pf} \itshape  theorem~\ref{error2}.  \upshape 
By proposition~\ref{spitzer}, we have
\begin{eqnarray*}
\E\left(M_t -M_t^n\right)&=&\int_0^t \frac{\E X_s^{+}}{s} ds- \sum_{k=1}^n \frac{\E X_{k\delta}^{+}}{k}.
\end{eqnarray*}
Define
\begin{eqnarray*}
h(s)=\frac{\E X_s^{+}}{s}, \quad s\in [0,t].
\end{eqnarray*}
In order to prove the theorem we need to show that $h$ is absolutely continuous 
(cf. Lemma~\ref{Imapp}). We will first show that the derivative (in the sense of distributions)
of $s\mapsto \E X_s^+$ is given by the function
\begin{eqnarray*}
\frac{d}{ds}\E (X_s)^{+}=\gamma_0 \P\left[X_s\geq0\right]+
\int_{\R}\E\left(\left(X_s+y\right)^{+}-\left(X_s\right)^{+}\right)\nu(dy),
\quad s\in (0,t).
\end{eqnarray*}
We first consider a  continuously differentiable function $f$  with bounded derivative. Since $X$ is a finite variation process,   It\^o's formula reduces to
\begin{eqnarray*}
f(X_s)=f(0)+\gamma_0\int_0^sf'\left(X_{{\tau}}\right)d\tau+\sum_{0\leq \tau\leq s}\left(f\left(X_{{\tau}}\right)-f\left(X_{{\tau^{-}}}\right)\right),
\end{eqnarray*}
so that 
\begin{eqnarray*}
\E f(X_s)=f(0)+\gamma_0\E\int_0^s f'\left(X_{{\tau}}\right)d\tau+\E\sum_{0\leq \tau\leq s}\left(f\left(X_{{\tau}}\right)-f\left(X_{{\tau^{-}}}\right)\right).
\end{eqnarray*}
The compensation formula (see \cite{bertoin}, preliminaries) yields that,  if 
\begin{equation}\label{comp}
\E\left[\int_0^sd\tau\int_{\R}\left|f\left(X_{{\tau}}+y\right)-f\left(X_{{\tau}}\right)\right|\nu(dy)\right]<\infty,
\end{equation}
 then
\begin{eqnarray*}
\E\sum_{0\leq \tau\leq s}\left(f\left(X_{{\tau}}\right)-f\left(X_{{\tau^{-}}}\right)\right)=\E\left[\int_0^sds\int_{\R}\left(f\left(X_{{\tau}}+y\right)-f\left(X_{{\tau}}\right)\right)\nu(dy)\right].
\end{eqnarray*}
Since $f$ is a Lipschitz function and $X$ is integrable, the condition \eqref{comp} is satisfied
and we have \small
\begin{eqnarray*}
\E f(X_s)&=&f(0)+\gamma_0\E\int_0^s f'\left(X_{{\tau}}\right)d\tau+\E\left[\int_0^sds\int_{\R}\left(f\left(X_{{\tau}}+y\right)-f\left(X_{{\tau}}\right)\right)\nu(dy)\right]
\\&=&f(0)+\E\left[\gamma_0\int_0^s f'\left(X_{{\tau}}\right)d\tau+\int_0^sds\int_{\R}\left(f\left(X_{{\tau}}+y\right)-f\left(X_{{\tau}}\right)\right)\nu(dy)\right].
\end{eqnarray*}\normalsize
Now, for  $\epsilon>0$, define
\begin{eqnarray*}
f_{\epsilon}(x)&=&\frac{x}{2}+\frac{\sqrt{\epsilon+x^2}}{2}, \quad x\in\R.
\end{eqnarray*}
Note  that $f_{\epsilon}$ is continuously differentiable and
\begin{eqnarray*}
f'_{\epsilon}(x)&=&\frac{1}{2}+\frac{x}{2\sqrt{\epsilon+x^2}}, \quad x\in\R,
\end{eqnarray*}
so that 
$
\left\|f'_{\epsilon}\right\|_{\infty}\leq1.
$
We can write \small
\begin{eqnarray*}
\E f_{\epsilon}(X_s)&=&\frac{1}{2}+\E\left[\gamma_0\int_0^s f'_{\epsilon}\left(X_{{\tau}}\right)d\tau+\int_0^sds\int_{\R}\left(f_{\epsilon}\left(X_{{\tau}}+y\right)-f_{\epsilon}\left(X_{{\tau}}\right)\right)\nu(dy)\right].
\end{eqnarray*}\normalsize
Note that the function $f_{\epsilon}$ converges uniformly to $x\rightarrow x^{+}$ when $\epsilon$ goes to $0$. And, for any $x\neq 0$,
\begin{eqnarray*}
\lim_{\epsilon\rightarrow0}f'_{\epsilon}(x)=\indicatrice_{x\geq0}.
\end{eqnarray*}
Moreover, for any $\tau>0$, $\P(X_{\tau}\neq0)=1$ (because $X$ have infinite activity), 
and, for any $x\in\R$,
\[
x^{+}\leq f_{\epsilon}(x)\leq\frac{x}{2}+\frac{\sqrt{\epsilon}+|x|}{2}
\leq x^{+}+\frac{\sqrt{\epsilon}}{2}.
\]
By dominated convergence, we get
\begin{eqnarray*}
\E (X_s)^{+}&=&\frac{1}{2}+\E\left[\gamma_0\int_0^s \indicatrice_{\left\{X_{{\tau}}\geq0\right\}}d\tau+\int_0^sds\int_{\R}\left(\left(X_{{\tau}}+y\right)^{+}-\left(X_{{\tau}}\right)\right)^{+}\nu(dy)\right]
\\&=&\frac{1}{2}+\gamma_0\int_0^s \P\left[X_{{\tau}}\geq0\right]d\tau+\int_0^sds\int_{\R}\E\left(\left(X_{{\tau}}+y\right)^{+}-\left(X_{{\tau}}\right)^{+}\right)\nu(dy).
\end{eqnarray*}
Hence
\begin{eqnarray*}
\frac{d}{ds}\E (X_s)^{+}&=&\gamma_0 \P\left[X_s\geq0\right]+
\int_{\R}\E\left(\left(X_s+y\right)^{+}-\left(X_s\right)^{+}\right)\nu(dy).
\end{eqnarray*}
Now, we have
\begin{eqnarray*}
&&h(s)-\int_{\R}y^{+}\nu(dy)=\frac{\E (X_s)^{+}}{s}-\int_{\R}y^{+}\nu(dy)
\\&=&\frac{1}{s}\int_0^s\left(\gamma_0\P\left[X_u\geq0\right]+ \int_{\R}\E\left(\left(X_u+y\right)^{+}-X_u^{+}\right)\nu(dy)\right)du-\int_{\R}y^{+}\nu(dy)
\\&=&\frac{\gamma_0}{s}\int_0^s\P\left[X_u\geq0\right]du+ \frac{1}{s}\int_0^s\int_{\R}\E\left(\left(X_u+y\right)^{+}-X_u^{+}-y^{+}\right)\nu(dy)du.
\end{eqnarray*}
But
\begin{eqnarray*}
\left(X_u+y\right)^{+}-X_u^{+}-y^{+}&=&\left(X_u+y\right)\indicatrice_{\{X_u+y>0\}}-X_u\indicatrice_{\{X_u>0\}}-y\indicatrice_{\{y>0\}}
\\&=&\left(X_u+y\right)\indicatrice_{\{X_u+y>0\}}-X_u\indicatrice_{\{X_u>0\}}-y\indicatrice_{\{y>0\}}
\\&=&X_u\left(\indicatrice_{\{X_u+y>0\}}-X_u\indicatrice_{\{X_u>0\}}\right)
\\&&-y\left(\indicatrice_{\{X_u+y>0\}}-X_u\indicatrice_{\{y>0\}}\right)
\\&=&-|X_u|\indicatrice_{\{yX_u<0,|y|>|X_u| \}}-|y|\indicatrice_{\{yX_u<0,|y|\leq|X_u|\} }
\\&=&-|X_u|\wedge|y|\indicatrice_{\{yX_u<0\}}.
\end{eqnarray*}
So \small
\begin{eqnarray*}
h(s)-\int_{\R}y^{+}\nu(dy)&=&\frac{\gamma_0}{s}\int_0^s\P\left[X_u\geq0\right]du- \frac{1}{s}\int_0^s\int_{\R}\E|X_u|\wedge|y|\indicatrice_{\{yX_u<0\}}\nu(dy)du.
\end{eqnarray*}\normalsize
It is now clear  that $h$ is continuous  on $(0,+\infty)$, and that its derivative is given by
\begin{eqnarray*}
h'(s)&=&u_s+v_s+w_s,
\end{eqnarray*}
where
\begin{eqnarray*}
u_s&=&\frac{\gamma_0}{s}\P\left[X_s\geq0\right]-
\frac{\gamma_0}{s^2}\int_0^s\P\left[X_u\geq0\right]du,\\
v_s&=&- \frac{1}{s}\int_{\R}\E\left[|X_s|\wedge|y|\indicatrice_{\{yX_s<0\}}\right]\nu(dy),\\
w_s&=&\frac{1}{s^2} \int_0^s\int_{\R}\E\left[|X_u|\wedge|y|\indicatrice_{\{yX_u<0\}}\right]\nu(dy)du.
\end{eqnarray*}
We will now show that
\begin{eqnarray*}
\int_{0}^t|h'(s)|ds<\infty
\end{eqnarray*}
We have $u_s=0$ if $\gamma_0=0$, and, for $\gamma_0\neq 0$, we can write
\begin{eqnarray*}
|u_s|&=&\left|\frac{\gamma_0}{s}\P\left[X_s\geq0\right]-\frac{\gamma_0}{s^2}\int_0^s\P\left[X_u\geq0\right]du\right|
\\&\leq&\frac{|\gamma_0|}{s}\int_0^1\left|\P\left[X_s\geq0\right]-\P\left[X_{su}\geq0\right]\right|du.
\end{eqnarray*}
Hence, by lemma~\ref{error2lemme2},
\begin{eqnarray*}
\int_0^t|u_s|ds<\infty.
\end{eqnarray*}
Besides, using the concavity of the function $x\in\R^{+}\rightarrow  x\wedge |y|$ and 
Proposition~\ref{levy_supremum_error}, we get
\begin{eqnarray*}
|v_s|&\leq&\frac{1}{s}\int_{\R}\E\left(|X_s|\wedge|y|\right)\indicatrice_{yX_s<0}\nu(dy)
\\&\leq&\frac{1}{s}\int_{\R}\E\left(|X_s|\wedge|y|\right)\nu(dy)
\\&\leq&\frac{1}{s}\int_{\R}\left(\E|X_s|\right)\wedge|y|\nu(dy)
\\&\leq&\frac{1}{s}\int_{\R}(cs)\wedge|y|\nu(dy),
\end{eqnarray*}
where the positive constant $c$ comes from Proposition~\ref{levy_supremum_error}.
Now, let $\hat{v}_s=\frac{1}{s}\int_{\R}(cs)\wedge|y|\nu(dy)$.
Using Fubini's theorem, we have
\begin{eqnarray*}
\int_0^t|\hat v_s|ds&=&\int_{\R}\nu(dy)\int_0^t\frac{ds}{s}(cs)\!\wedge|y|\\
&\leq& c\int_{\R}\int_0^{\frac{|y|}{c}}ds\nu(dy)+\int_{\R}\int_{\frac{|y|}{c}}^t\frac{1}{s}|y|\indicatrice_{\{|y|\leq ct\}}ds\nu(dy)
\\&=& \int_{\R}|y|\nu(dy) +\int_{\R}\log\left(\frac{ct}{|y|}\right)|y|\indicatrice_{\{|y|\leq ct\}}\nu(dy)
\\&=& \int_{\R}|y|\nu(dy) +\int_{|y|\leq ct}\log\left(\frac{ct}{|y|}\right)|y|\nu(dy)
\\&<&\infty.
\end{eqnarray*}
Note that the last integral is finite, due to the assumption on the L\'evy measure.
For the term $w_s$, we have
\begin{eqnarray*}
|w_s|&\leq&\frac{1}{s^2} \int_0^s\int_{\R}(cu)\wedge|y|\nu(dy)du
\\&\leq&\frac{1}{s^2} \int_0^s\int_{\R}(cs)\wedge|y|\nu(dy)du
\\&=&\frac{1}{s}\int_{\R}(cs)\wedge|y|\nu(dy)=\hat v_s.
\end{eqnarray*}
We deduce that
\begin{eqnarray*}
\int_0^t|w_s|ds&<&\infty.
\end{eqnarray*}
Therefore, we have proved that $h$ is absolutely continuous. 
Using lemma~\ref{Imapp} and theorem~\ref{spitzer} we complete the proof.
\end{pf}

\section{Extension of the Asmussen-Glynn-Pitman Theorem}
\label{sec:asmussentheorem}
The continuity correction results of Broadie Glasserman and Kou for lookback options
within the Black-Scholes model are based on a result due to Asmussen, Glynn and Pitman,
about the weak convergence of the normalized difference between the continuous and discrete 
maximum of Brownian motion(see \cite{asmussen-al95}, Theorem 1). In this section, we extend this result
to L\'evy processes with finite activity and a non-trivial Brownian component,
 i.e. a L\'evy process with generating
triplet $(\gamma,\sigma^2, \nu)$, where $\sigma^2>0$ and $\nu$ is a finite measure.

The following statement is a reformulation of the Asmussen-Glynn-Pitman Theorem.
It can be deduced from a careful reading of the proof of Theorem 1 in \cite{asmussen-al95} (see particularly pages 879 to 883,
and Remark 2).
\begin{thm}\label{Thm-AGP}
  Consider four real numbers $a$, $b$, $x$ and $y$, with $0\leq a<b$. Let $\beta =(\beta_t)_{a\leq t\leq b}$
  be a Browian bridge from $x$ to $y$ over the time interval $[a,b]$ (so that $\beta_a=x$ and $\beta_b=y$) and let $t$ be a fixed positive number.
  Denote by $M$ the supremum of $\beta$ and, for any positive integer $n$, by $M^n$ the discrete supremum
  associated with a mesh of size $\frac{t}{n}$, so that
  \[
  M=\sup_{a\leq t\leq b}\beta_t\quad\mbox{and}\quad M^n=\sup_{k\in I_n}\beta_{\frac{kt}{n}},
  \mbox{ where } I_n=\left\{k\in \N\;|\; \frac{kt}{n}\in [a,b]\right\}.
  \]
  Then, as $n$ goes to infinity, the pair $\left(\sqrt{n}\left(M-M^n\right), \beta\right)$ converges in distribution to the pair $(\sqrt{t}W,\beta)$
  where $W$ is independent of $\beta$ and can be written as
  \begin{equation}\label{W}
  W=\min_{\{j\in\Z\}}\check{R}(U+j).
  \end{equation}
  Here $(\check {R}(t))_{t\in \R}$ is a two sided three dimensional Bessel process (i.e. $\check {R}(t)=R_1(t)$
  for $t\geq 0$ and $\check R(t)=R_2(-t)$ for $t<0$, where $R_1$ and $R_2$ are independent copies of the usual three dimensional
   Bessel process, starting from $0$) and $U$ is uniformly distributed on $[0,1]$ and independent of $\check{R}$.
\end{thm}
We can now state and prove the main result of this section.
\begin{thm}\label{supremum_error_law_convergence}
Let $X=(X_t)_{t\geq 0}$ be a finite activity L\'evy process with generating triplet $(\gamma, \sigma^2, \nu)$
satisfying $\sigma^2>0$. For a fixed positive real number $t$, consider the continuous supremum of $X$
over $[0,t]$ and, for any positive integer $n$, the discrete supremum associated with a mesh of size $\frac{t}{n}$, that is
\[
M_t=\sup_{0\leq s\leq t}X_s \quad\mbox{and}\quad
M^n_t=\sup_{k=0,1,\ldots,n}X_{\frac{kt}{n}}.
\]
Then, as $n$ goes to infinity, the pair $\left(\sqrt{n}\left(M_t-M^n_t\right), X^{(t)}=(X_s)_{0\leq s\leq t}\right)$ converges in distribution to the pair $(\sigma\sqrt{t}W,X^{(t)})$
  where $W$ is independent of $X^{(t)}$ and given by (\ref{W}).
 \end{thm}
 Note that, in the above statement, $X^{(t)}$ is viewed as a random variable with values in the space of c\`ad-l\`ag functions
 defined on the interval $[0,t]$, which can be endowed with the Skorohod topology.

\begin{pf} \itshape theorem~\ref{supremum_error_law_convergence}. \upshape
We will prove that for any bounded and continuous function $f$ and for any bounded random variable $Z$
which is measurable with respect to the $\sigma$-algebra generated by the random variables $X_s$, $0\leq s\leq t$,
we have
\begin{equation}\label{2}
\lim_{n\to \infty}\E\left(f\left(\sqrt{n}(M_t-M^n_t)\right)Z\right)=\E\left(f\left(\sigma\sqrt{t}W\right)\right)\E(Z).
\end{equation}
Since $X$ is a finite activity process, it admits the following representation
\[
X_s=\gamma_0 s+\sigma B_s+\sum_{j=1}^{N_s}Y_j,\quad s\geq 0,
\]
where $B$ is a standard Brownian motion, $N$ is a Poisson process with intensity $\lambda=\nu(\R)$,
and the random variables $Y_j$ are iid with distribution $\frac{\nu}{\nu(\R)}$. Note that $B$, $N$ and the $Y_j$'s are independent.

By conditioning with respect to $N_t$, we have
\[
\E\left(f\left(\sqrt{n}(M_t-M^n_t)\right)Z\right)=\sum_{m=0}^\infty\E\left(f\left(\sqrt{n}(M_t-M^n_t)\right)Z\;|\;N_t=m\right)\P(N_t=m).
\]
Note that, conditionally on $\{N_t=0,X_t=y\}$, the process $\frac{X^{(t)}}{\sigma}$ is a Brownian bridge from $0$ to $\frac{y}{\sigma}$ so that, using Theorem~\ref{Thm-AGP},
\[
\lim_{n\rightarrow+\infty}\E\left(f\left(\sqrt{n}(M_t-M^n_t)\right)Z\;|\;N_t=0\right)=\E\left(f\left(\sigma\sqrt{t}W\right)\right)\E\left(Z\;|\;N_t=0\right).
\]
For the conditional expectation given $\{N_t=m\}$, $m\geq 1$, we condition further with respect to the
jump times, to the values of $X$ and to the values of the left-hand limits at the jump times. Denote by
$T_1$, $T_2$,\ldots, $T_j$,\ldots  the jump times of the Poisson process $N$.
For any numbers $0<t_1<t_2<\ldots<t_m<t$, $x_1$,\ldots, $x_m$, $y_1$,\ldots,$y_m$, $y_{m+1}$, let
\[
A_m=\left\{N_t=m,T_i=t_i, X_{T_i^-}=x_i,X_{T_i}=y_i, i=1,\ldots, m, X_t=y_{m+1}\right\}.
\]
We observe that, conditionally  on $A_m$, the random processes $\beta^0$, \ldots, $\beta^m$
defined by
\[
\beta^j_s=\left\{
\begin{array}{l}\displaystyle
\frac{1}{\sigma}X_s\;\mbox{ if } s\in [t_j,t_{j+1}),\\
\\
\displaystyle
\frac{1}{\sigma}X_{t_{j+1}^-} \;\mbox{ if } s=t_{j+1},
\end{array}
           \right.
\]
with $t_0=0$ and $t_{m+1}=t$, are independent Brownian bridges over the intervals $[t_j,t_{j+1}]$.
Introduce the random variables
\[
M^j=\sup_{t_j\leq s\leq t_{j+1}}\beta^j_s, \quad
M^{j,n}=\sup_{k\in I^j_n}\beta^j_{\frac{kt}{n}},
\]
where $I^j_n=\left\{k\in \N\;|\; t_j\leq \frac{kt}{n}\leq t_{j+1}\right\}$. Conditionally on $A_m$, the random variables
$M^j$ are independent and each of them admits a density. Therefore, with probability one, one of them has to be strictly 
larger than the others. For $j=0$,\ldots, $m$, set
\[
A^j_m=\{M^j>M^i \mbox{ for}\; i\neq j\}.
\]
Conditionally on $A_m$, we have
\[
f\left(\sqrt{n}(M_t-M^n_t)\right)Z=\sum_{j=0}^m\ind{A^j_m}f\left(\sqrt{n}(\sigma M^j-M^n_t)\right)G_j(\beta^0,\ldots,\beta^m),
\]
for some bounded Borel functions $G_j$ defined on the space $\prod_{j=0}^mC([t_j,t_{j+1}])$.
Now, on the set $A^j_m$, we have, for $n$ large enough, $M^n_t=\sigma M^{j,n}$. This follows
from the fact that the maximum of $\beta^j$ is attained at an  interior point of the interval $(t_j,t_{j+1})$
and the fact that for $n$ large enough, some elements of $I^j_n$ are arbitrarily close to this point.
Therefore, for $n$ large enough, we have
\[
f\left(\sqrt{n}(M_t-M^n_t)\right)Z=\sum_{j=0}^m\ind{A^j_m}f\left(\sigma \epsilon^j_n\right)G_j(\beta^0,\ldots,\beta^m),
\]
with $\epsilon^j_n=\sqrt{n} (M^j-M^{j,n}_t)$.
We deduce from Theorem~\ref{Thm-AGP} and the independence of the Brownian bridges that \small
\begin{eqnarray*}
\lim_{n\to\infty}\E\left(f\left(\sqrt{n}(M_t-M^n_t)\right)Z\;|\;A_m\right)&=&
     \sum_{j=0}^m\lim_{n\to\infty}\E\left(\ind{A^j_m}f\left(\sigma\epsilon^j_n\right)G_j(\beta^0,\ldots,\beta^m)\;|\; A_m\right)\\
     &=&\sum_{j=0}^m\E\left(f\left(\sigma\sqrt{t} W\right)\right)
     \E\left(\ind{A^j_m}G_j(\beta^0,\ldots,\beta^m)\;|\; A_m\right)\\
     &=&\E\left(f\left(\sigma\sqrt{t} W\right)\right)\E(Z\;|\; A_m).
\end{eqnarray*} \normalsize
Hence, for all $m\geq 1$,
\[
\lim_{n\to\infty}\E\left(f\left(\sqrt{n}(M_t-M^n_t)\right)Z\;|\;N_t=m\right)=\E\left(f\left(\sigma\sqrt{t} W\right)\right)\E(Z\;|\; N_t=m),
\]
so that (\ref{2}) follows easily.
\end{pf}

In order to use the convergence in distribution above, we sometimes need to switch 
between limit and expected value. For that purpose, the following result of uniform integrability will be useful.

\begin{lemme}\label{uniform_integrability}
Let $X$ be a finite activity L\'evy process with generating triplet $(\gamma,\sigma^2,\nu)$, satisfying $\sigma>0$. Fix $t>0$ and set $\epsilon_n=M_t-M_t^n$. Then the sequence $(\sqrt{n}\epsilon_ne^{-M_t})_{n\geq1}$ is uniformly integrable. If in addition $\E e^{qM_t}<\infty$ for some $q>2$, then the sequence $(\sqrt{n}\epsilon_ne^{M_t})_{n\geq1}$ is uniformly integrable. 
\end{lemme}

\begin{pf} \itshape lemma~\ref{uniform_integrability}. \upshape
 We will prove that $(\sqrt{n}\epsilon_ne^{M_t})_{n\geq1}$ is uniformly integrable. The other case can be easily deduced. We will use the same notations as in the proof of 
 Theorem~\ref{supremum_error_law_convergence}. 
 Note that on the set $\{N_t=0\}$, we have 
 $X_s=\gamma_0 s+\sigma B_s$ for $0\leq s\leq t$, so that
 the uniform integrability of the sequence 
 $(\sqrt{n}\epsilon_ne^{M_t}\ind{\{N_t=0\}})_{n\geq1}$
 follows from Lemma 6 in \cite{asmussen-al95}.
 On the event $\{N_t\geq 1\}$, we will need to rule out the case when
 there is no jump between two mesh-points. So, 
 we introduce the event
 \[
 \Lambda_n=\{N_t\geq 1 \mbox{ and } \exists j\in\{1,\ldots,N_t\}\;
       T_j-T_{j-1}\leq t/n\}\cup\{t-T_{N_t}\leq t/n\}.
\]
Note that
\begin{eqnarray*}
\P(\Lambda_n)&\leq & \P(t-T_{N_t}\leq t/n)
                                 +\E \sum_{j=1}^{N_t}\ind{\{T_j-T_{j-1}\leq t/n\}}\\
                        &\leq &\E N_t(N_t+1)/n,
\end{eqnarray*}
where we have used the inequalities 
$\P(t-T_{N_t}\leq t/n\;|\; N_t=l)\leq   l/n$ and
$\P( T_j-T_{j-1}\leq t/n \;|\; N_t=l)\leq   l/n$ (cf. \cite{dia}, 
Proposition~5.5).
 
 Therefore, we have, using $\epsilon_n\leq M_t$ and
 H\"older's inequality,
 \begin{eqnarray*}
\E \left(\sqrt{n}\epsilon_ne^{M_t}\ind{\Lambda_n}\right)
     &\leq &
     \sqrt{n}\left(\E M_t^pe^{pM_t}\right)^{\frac{1}{p}}\left(\P(\Lambda_n)\right)^{1-\frac{1}{p}},
\end{eqnarray*}
for every $p>1$. Since $\E e^{q M_t}<\infty$ for some $q>2$,
 we can choose $p>2$. Hence
 \[
 \lim_{n\to \infty}\E\left( \sqrt{n}\epsilon_ne^{M_t}\ind{\Lambda_n}
    \right)
     =0.
 \]
Now, we want to prove that the sequence 
$(\sqrt{n}\epsilon_ne^{M_t}
\ind{\{N_t\geq 1\}\cap\Lambda_n^c})_{n\geq1}$ is uniformly
integrable.
 
Fix $m\geq 1$ and $t_1$,\ldots, $t_m$ satisfying 
$0<t_1<\ldots<t_m<t$. Conditionaly  on 
$\{N_t=m,T_1=t_1,\ldots,T_m=t_m\}\cap \Lambda_n^c$, 
we have, with probability one,
\[
\epsilon_n=\sum_{j=0}^m\left(M^j-M^n_t\right)
      \ind{\{M^j>\max_{i\neq j} M^i\}},
\]
where $M^j=\sup_{t_j\leq s<t_{j+1}}X_s$, $t_0=0$ and $t_{m+1}=t$.
Moreover, due to the definition of $\Lambda_n$, 
each subinterval $[t_j,t_{j+1})$ contains 
at least one mesh point. Denote 
\begin{eqnarray*}
k_j&=&\min\{k\in\{0,1,\ldots,n\}\;|\; kt/n\geq t_j\}\\
l_j&=&\max\{k\in\{0,1,\ldots,n\}\;|\; kt/n\leq  t_{j+1}\},
\end{eqnarray*}
and let $s^*$ be a point at which the supremum of $X_s$
over $[t_j,t_{j+1})$ is attained. If $s^*\in (t_j,k_jt/n)$, we can write
$M^j-M^n_t\leq \sup_{s\in (t_j,k_jt/n)}(X_s-X_{k_jt/n})$.
If $s^*\in (l_jt/n,t_{j+1})$, we have 
$M^j-M^n_t\leq \sup_{s\in (l_jt/n,t_{j+1})}(X_s-X_{l_jt/n})$.
Hence
\[
M^j-M^n_t\leq \delta_{n,j}+\epsilon_{n,j}+\eta_{n,j},
\]
 where
 \[
 \delta_{n,j}=\sup_{s\in (t_j,k_jt/n)}(X_s-X_{k_jt/n}),\quad
 \eta_{n,j}=\sup_{s\in (l_jt/n,t_{j+1})}(X_s-X_{l_jt/n}),
  \]
 and
 \[
 \epsilon_{n,j}=\sup_{k_jt/n\leq s\leq l_jt/n}X_s-
            \max_{k_j\leq k\leq l_j}X_{kt/n}.
 \]
Observe that
\begin{eqnarray}
\delta_{n,j}&=&\sup_{s\in (t_j,k_jt/n)}
  \left[\gamma_0s+\sigma B_s-\left(\gamma_0 \frac{k_jt}{n}
         +\sigma B_{k_jt/ n}\right)\right]\nonumber\\
         &\leq &
            |\gamma_0|\frac{t}{n}+\sigma\sup_{s\in (t_j,k_jt/n)}
                 \left|B_s-B_{k_jt/n}\right|.\label{delta}
\end{eqnarray}
 Similarly,
 \begin{eqnarray}\label{eta}
\eta_{n,j}&\leq &
            |\gamma_0|\frac{t}{n}+\sigma\sup_{s\in (l_jt/n,t_{j+1})}
                 \left|B_s-B_{l_jt/n}\right|.
\end{eqnarray}
Note that $|t_j-k_jt/n|\leq t/n$ and $t_{j+1}-l_jt/n\leq t/n$.
Therefore, we easily deduce from (\ref{delta}) (resp. (\ref{eta}))
that the conditional expectation of any power of $\sqrt{n}\delta_{n,j}$
(resp. $\sqrt{n}\eta_{n,j}$) is bounded by a constant which is independent of
the conditioning.
We also have
 \begin{eqnarray*}
\epsilon_{n,j}&=&
           \sup_{0\leq s\leq (l_j-k_j)t/n}\beta^j_s-
            \max_{0\leq k\leq l_j-k_j}\beta^j_{kt/n},
\end{eqnarray*}
where $\beta^j_s=\gamma_0 s+\sigma(B_{s+k_jt/n}-B_{k_jt/n})$.
Using Lemma 6 of \cite{asmussen-al95}, we see that
the  conditional expectation of any power of $\sqrt{n}\epsilon_{n,j}$
 is bounded by a constant which is independent of
the conditioning. We conclude from this discussion that, for any $p>1$,
\[
\E\left[
\left(\sqrt{n}\epsilon_n\ind{\Lambda_n^c\cap\{N_t\geq 1\}}\right)^p
    \;|\; N_t\right]
     \leq 
       C_pN_t^p,
\]
 where $C_p$ is a deterministic constant which depends only on 
 $p$, $\gamma_0$, $\sigma$ and $t$. The uniform integrability
 of $\sqrt{n}\epsilon_ne^{M_t}$ follows easily.
\end{pf}

\section{Continuity correction}
\label{sec:contcorr}
In this section, we extend the results of Broadie-Glasserman-Kou (1999)  
on lookback and hindsight options to the jump-diffusion model. Let $(S_t)_{t\in[0,T]}$ be the price of a 
security modeled  as a stochastic process on a filtered probability space 
$\left(\Omega,\F,(\F_t)_{t\in[0,T]},\P\right)$. 
The $\sigma$-algebra $\F_t$ represents the historical information on the price until time $t$. 
Under the exponential L\'evy model, the process $S$ behaves as the exponential of a L\'evy process
\begin{eqnarray*}
S_t=S_0e^{X_t},
\end{eqnarray*}
where $X$ is a L\'evy process with generating triplet $(\gamma,\sigma^2,\nu)$. 
The considered probability is a risk-neutral probability, under which the process 
\\$\left(e^{-(r-\delta)t}S_t\right)_{t\in[0,T]}$ is a martingale. 
The parameter $r$ is the risk-free interest rate, and $\delta$ is the dividend rate.
 The options we will consider in the sequel will have as underlying the asset with price $S$. 
 We will denote by $K$ the strike price of the option (in the case of hindsight options). 
Figure \ref{payoffstab} gives the payoffs of lookback and hindsight options. 
The corresponding prices are the expected values of the discounted payoffs.

\begin{figure}[h]
\begin {center}
\begin{tabular}{|c|c|c|}
 \hline 
 Option &continuous &discrete\\ \hline
 Lookback call & $S_T-S_0e^{m_T}$ &$S_T-S_0e^{m_T^n}$\\ \hline
 Lookback  put&$S_0e^{M_T}-S_T$&$S_0e^{M_T^n}-S_T$\\ \hline
 Hindsight call & $\left(S_0e^{M_T}-K\right)^{+}$&$\left(S_0e^{M_T^n}-K\right)^{+}$ \\ \hline
 Hindsight put&$\left(K-S_0e^{m_T}\right)^{+}$&$\left(K-S_0e^{m_T^n}\right)^{+}$ \\ \hline
\end{tabular}
\caption{The payoffs of lookback and hindsight options.}
\label{payoffstab}
\end {center}
\end{figure}

The r.v. $m_T$ and $m_T^n$ in table~\ref{payoffstab} satisfy
\begin{eqnarray*}
m_T=\inf_{0\leq s\leq T}X_{s}, \ m_T^{n}=\min_{0\leq k\leq n}X_{k\Delta t},
\end{eqnarray*}
where $\Delta t=\frac{T}{n}$. The results we are going to show depend on the assumptions made on the process $X$. That is why we need to introduce the following assumptions:

\begin{description}
 \item[H1] \normalsize \mdseries  $X$ is an integrable L\'evy process with finite activity, satisfying $\sigma>0$ and there exists $q>2$ such that $\E e^{qM_T}<\infty$; 
\item[H2] \normalsize \mdseries  $X$ is an integrable L\'evy process with finite activity, satisfying $\sigma>0$.
\end{description}
Let $W$ be the r.v. defined in theorem~\ref{Thm-AGP}. We set $\beta_1=\E W=-\frac{\zeta\left(\frac{1}{2}\right)}{\sqrt{2\pi}}$, where $\zeta$ is the Riemann zeta function.

At a given time $t\in[0,T)$, the value of the continuous lookback  put is given by
\begin{eqnarray*}
V\left(S_{+}\right)=e^{-r(T-t)}\E \max\left(S_{+},\max_{t\leq u\leq T}S_u\right)-S_te^{-\delta (T-T)},
\end{eqnarray*}
where $S_{+}=\max_{0\leq u\leq t}S_u$  is the predetermined maximum. The continuous value 
of the lookback call will depend similarly on $S_{-}=\min_{0\leq u\leq t}S_u$ 
(the predetermined minimum) and on $\min_{t\leq u\leq T}S_u$.
The price of the discrete lookback  put at the $k$-{th} fixing date is given by
\begin{eqnarray*}
V_n\left(S_{+}\right)=e^{-r\Delta(n-k)}\E \max\left(S_{+},\max_{k\leq j\leq n}S_{j\Delta t}\right)-S_{k\Delta t}e^{-\delta (n-k)\Delta t},
\end{eqnarray*}
where $S_{+}=\max_{0\leq j\leq k}S_{j\Delta t}$. The discrete call value will depend similarly on $S_{-}=\min_{0\leq j\leq k}S_{j\Delta t}$ and on $\min_{k\leq j\leq n}S_{j\Delta t}$. 

\begin{prop}\label{lbt}
The price of a discrete lookback option at the $k$-{th} fixing date and the price of the continuous lookback option at $k\Delta t$ satisfy \small
\begin{eqnarray*}
&&V_n\left(S_{\pm}\right)=e^{\mp\beta_1\sigma\sqrt{\frac{T}{n}}}V\left(S_{\pm}e^{\pm\beta_1\sigma\sqrt{\frac{T}{n}}}\right)\pm\left(e^{\mp\beta_1\sigma\sqrt{\frac{T}{n}}}-1\right)e^{-\delta(T-t)}S_t + o\left(\frac{1}{\sqrt{n}}\right)\\
&&V\left(S_{\pm}\right)=e^{\pm\beta_1\sigma\sqrt{\frac{T}{n}}}V_n\left(S_{\pm}e^{\mp\beta_1\sigma\sqrt{\frac{T}{n}}}\right)\pm\left(e^{\pm\beta_1\sigma\sqrt{\frac{T}{n}}}-1\right)e^{-\delta(T-t)}S_t + o\left(\frac{1}{\sqrt{n}}\right),
\end{eqnarray*}\normalsize
where in $\pm$ and $\mp$, the top case applies for puts and the bottom case for calls. The relations for the put are true under $H1$, and those for the call under $H2$.
\end{prop}
These formulas  are the same as those  found by Broadie, Glasserman and Kou (1999) 
for the Black-Scholes model. 

\begin{pf} \itshape proposition~\ref{lbt}. \upshape
Since we have theorem~\ref{supremum_error_law_convergence} and lemma~\ref{uniform_integrability}, the proofs of the above proposition is similar to the proof of theorem $3$ of \cite{broadie-glasserman-kou99}. For example to relate discrete lookback put with respect to continuous lookback put, we need to prove that for $x\in\R$
\begin{eqnarray*}
\E \left(e^{M_T^n}-x\right)^{+}=e^{-\beta_1\sigma\sqrt{\frac{T}{n}}}\E \left(e^{M_T}-e^{\beta_1\sigma\sqrt{\frac{T}{n}}}x\right)^{+}+o\left(\frac{1}{\sqrt{n}}\right).
\end{eqnarray*}
In fact we have to show first that
\begin{eqnarray*}
\E \left(e^{M_T}-x\right)^{+}&=&\E \left(e^{M_T}-e^{M_T^n}\right)\indicatrice_{\left\{e^{M_T}>x\right\}}+\E \left(e^{M_T^n}-x\right)^{+}\\&&+\E \left(e^{M_T^n}-x\right)\indicatrice_{\left\{e^{M_T^n}\leq x<e^{M_T}\right\}}.
\end{eqnarray*}
So
\begin{eqnarray*}\label{exp+1}
\E \left(e^{M_T^n}-x\right)^{+}&=&\E \left(e^{M_T}-x\right)^{+}-\E \left(e^{M_T}-e^{M_T^n}\right)\indicatrice_{\left\{e^{M_T}>x\right\}}\\&&-\E \left(e^{M_T^n}-x\right)\indicatrice_{\left\{e^{M_T^n}\leq x<e^{M_T}\right\}}.
\end{eqnarray*}
But
\begin{eqnarray*}
\E \left|e^{M_T^n}-x\right|\indicatrice_{\left\{e^{M_T^n}\leq x<e^{M_T}\right\}}&\leq&\E \left(e^{M_T}-e^{M_T^n}\right)\indicatrice_{\left\{e^{M_T^n}\leq x<e^{M_T}\right\}}\\
&\leq&\E \left(M_T-M_T^n\right)e^{M_T}\indicatrice_{\left\{e^{M_T^n}\leq x<e^{M_T}\right\}}.
\end{eqnarray*}
Moreover the sequence 
\begin{eqnarray*}
\left(\sqrt{n}\left(M_T-M_T^n\right)e^{M_T}\indicatrice_{\left\{e^{M_T^n}\leq x<e^{M_T}\right\}}\right)_{n\geq1}
\end{eqnarray*}
is uniformly integrable (by lemma~\ref{uniform_integrability}). So
\begin{eqnarray*}
\lim_{n\rightarrow +\infty}\E \sqrt{n}\left(M_T-M_T^n\right)e^{M_T}\indicatrice_{\left\{e^{M_T^n}\leq x<e^{M_T}\right\}}
&=& 0.
\end{eqnarray*}
On the other hand, using theorem~\ref{supremum_error_law_convergence} and lemma~\ref{uniform_integrability}, we get
\begin{eqnarray*}
E \left(e^{M_T}-e^{M_T^n}\right)\indicatrice_{\left\{e^{M_T}>x\right\}}
&=&\sigma\beta_1\sqrt{\frac{T}{n}}\E e^{M_T}\indicatrice_{\left\{e^{M_T}>x\right\}}+o\left(\frac{1}{\sqrt{n}}\right).
\end{eqnarray*}
Thus
\begin{eqnarray*}
\E \left(e^{M_T^n}-x\right)^{+}&=&\E \left(e^{M_T}-x\right)^{+}-\sigma\beta_1\sqrt{\frac{T}{n}}\E e^{M_T}\indicatrice_{\left\{e^{M_T}>x\right\}}+o\left(\frac{1}{\sqrt{n}}\right)\\
&=&e^{-\sigma\beta_1\sqrt{\frac{T}{n}}}\E\left(e^{M_T}-xe^{\sigma\beta_1\sqrt{\frac{T}{n}}}\right)\indicatrice_{\left\{e^{M_T}>x\right\}} +o\left(\frac{1}{\sqrt{n}}\right)\\
&=&e^{-\sigma\beta_1\sqrt{\frac{T}{n}}}\E\left(e^{M_T}-xe^{\sigma\beta_1\sqrt{\frac{T}{n}}}\right)\indicatrice_{\left\{x<e^{M_T}\leq xe^{\sigma\beta_1\sqrt{\frac{T}{n}}}\right\}}
\\&&+ e^{-\sigma\beta_1\sqrt{\frac{T}{n}}}\E\left(e^{M_T}-xe^{\sigma\beta_1\sqrt{\frac{T}{n}}}\right)^{+} +o\left(\frac{1}{\sqrt{n}}\right).
\end{eqnarray*}
But, we can show that
\begin{eqnarray*}
\E\left(e^{M_T}-xe^{\sigma\beta_1\sqrt{\frac{T}{n}}}\right)\indicatrice_{\left\{x<e^{M_T}\leq xe^{\sigma\beta_1\sqrt{\frac{T}{n}}}\right\}}&=&o\left(\frac{1}{\sqrt{n}}\right).
\end{eqnarray*}
Hence
\begin{eqnarray*}
\E \left(e^{M_T^n}-x\right)^{+}=e^{-\sigma\beta_1\sqrt{\frac{T}{n}}} \E\left(e^{M_T}-xe^{\sigma\beta_1\sqrt{\frac{T}{n}}}\right)^{+}+o\left(\frac{1}{\sqrt{n}}\right).
\end{eqnarray*}
The others cases can be derived in the same way. Detailed proofs are given in \cite{dia}.
\end{pf}

For hindsight options, we have similar results as for the lookback case.
 The price of a continuous hindsight  call option at time $t$ with a predetermined maximum 
 $S_{+}$ and strike $K$ is
\begin{eqnarray*}
V\left(S_{+},K\right)=e^{-r(T-t)}\E \left(\max\left(S_{+},\max_{t\leq u\leq T}S_u\right)-K\right)^{+}.
\end{eqnarray*}
Similarly,  for the put we have
\begin{eqnarray*}
V\left(S_{-},K\right)=e^{-r(T-t)}\E \left(K-\min\left(S_{-},\min_{t\leq u\leq T}S_u\right)\right)^{+}.
\end{eqnarray*}
The discrete versions at the $k$-{th} fixing date are
\begin{eqnarray*}
V_n\left(S_{+},K\right)=e^{-r\Delta t(n-k)}\E \left(\max\left(S_{+},\max_{k\leq j\leq n}S_{j\Delta t}\right)-K\right)^{+}
\end{eqnarray*}
and
\begin{eqnarray*}
V_n\left(S_{-},K\right)=e^{-r\Delta t(n-k)}\E \left(K-\min\left(S_{-},\min_{k\leq j\leq n}S_{j\Delta t}\right)\right)^{+}.
\end{eqnarray*}

\begin{prop}\label{hst}
The prices of a discrete hindsight option at the $k$-{th} fixing date and its continuous version at $k\Delta t$, satisfy
\[
V_n\left(S_{\pm},K\right)=e^{\mp\beta_1\sigma\sqrt{\frac{T}{n}}}V\left(S_{\pm}e^{\pm\beta_1\sigma\sqrt{\frac{T}{n}}},Ke^{\pm\beta_1\sigma\sqrt{\frac{T}{n}}}\right)+ o\left(\frac{1}{\sqrt{n}}\right)
\]
and
\[
V\left(S_{\pm},K\right)=e^{\pm\beta_1\sigma\sqrt{\frac{T}{n}}}V_n\left(S_{\pm}e^{\mp\beta_1\sigma\sqrt{\frac{T}{n}}},Ke^{\mp\beta_1\sigma\sqrt{\frac{T}{n}}}\right)+ o\left(\frac{1}{\sqrt{n}}\right),
\]
where in $\pm$ and $\mp$, the top case applies for calls and the bottom for puts. The relations for the calls are true under $H1$, and those for the put under $H2$.
\end{prop}
To explain the above proposition one can say that, in order to price a continuous (resp. discrete) hindsight option using a discrete (resp. continuous) one, we must shift the predetermined extremum and the strike. Proposition~\ref{hst} can be deduced from proposition~\ref{lbt}, thanks to the relations between lookback and hindsight options.

\begin{rmq}\label{lbt2}\rm
If the process $X$ is an integrable L\'evy process with generating triplet $(\gamma,0,\nu)$, satisfying $\nu(\R)<\infty$, then the price of a discrete lookback option and its continuous version at time $k\Delta t$ satisfy
\begin{enumerate}
\item for the call
\begin{eqnarray*}
&&V_n\left(S_{-}\right)=V\left(S_{-}\right)+\frac{\alpha}{n}+o\left(\frac{1}{n}\right),
\end{eqnarray*}
where the constant $\alpha$ can be derived explicitly,
\item for the put, if there exists $\beta>1$ such that $\E e^{\beta M_T}<\infty$, then
\begin{eqnarray*}
V_n\left(S_{+}\right)=V\left(S_{+}\right)+o\left(\frac{1}{n^{\frac{\beta-1}{\beta}}}\right).
\end{eqnarray*}
\end{enumerate}
\end{rmq}

The proof of these results can be found in \cite{dia}.

\section{Upper bounds}
\label{sec:bounds}

In the infinite activity case and if there is no Brownian part, the prices of the discrete and continuous calls are close to each other. The following proposition is a consequence of theorems \ref{error} and \ref{error2}.
\begin{prop}\label{lbct3}
Suppose that $X$ is an integrable infinite activity L\'evy process with generating triplet $(\gamma,0,\nu)$. Then the prices of a discrete call option at the $k^{th}$ fixing date and its continuous version at 
$k\Delta t$ satisfy
\begin{enumerate}
\item
\begin{eqnarray*}
&&V_n\left(S_{-}\right)=V\left(S_{-}\right)+  o\left(\frac{1}{\sqrt{n}}\right).
\end{eqnarray*}
\item If $\int_{|x|\leq1}|x|\nu(dx)<\infty$,
\begin{eqnarray*}
&&V_n\left(S_{-}\right)=V\left(S_{-}\right)+  O\left(\frac{\log(n)}{n}\right).
\end{eqnarray*}
\item If $\int_{|x|\leq1}|x|\log(|x|)\nu(dx)<\infty$,
\begin{eqnarray*}
&&V_n\left(S_{-}\right)=V\left(S_{-}\right)+ O\left(\frac{1}{n}\right).
\end{eqnarray*}
\end{enumerate}
\end{prop}

In the put case, the error between continuous and discrete prices depends on the integrability of the exponential of the supremum of the L\'evy process driving the underlying asset.
\begin{thm}\label{lbpt3}
Suppose that $X$ is an infinite activity L\'evy process with generating triplet $(\gamma,0,\nu)$ and there exists $\beta>1$ such that $\E e^{\beta M_T}<\infty$. Then the price of a discrete put option at the $k$-{th} fixing date and its continuous version at $k\Delta t$, satisfy
\begin{enumerate}
\item We have, for any $\epsilon>0$,
\begin{eqnarray*}
&&V_n\left(S_{+}\right)=V\left(S_{+}\right)+ O\left(\frac{1}{n^{\frac{\beta-1}{2\beta}-\epsilon}}\right).
\end{eqnarray*}
\item If $\int_{|x|\leq1}|x|\nu(dx)<\infty$, we have, for any $\epsilon>0$,
\begin{eqnarray*}
&&V_n\left(S_{+}\right)=V\left(S_{+}\right)+O\left(\left(\frac{\log(n)}{n}\right)^{\frac{\beta-1}{\beta}-\epsilon}\right).
\end{eqnarray*}
\item If $\int_{|x|\leq1}|x|\log(|x|)\nu(dx)<\infty$, we have, for any $\epsilon>0$,
\begin{eqnarray*}
&&V_n\left(S_{+}\right)=V\left(S_{+}\right)+O\left(\frac{1}{n^{\frac{\beta-1}{\beta}-\epsilon}}\right).
\end{eqnarray*}
\end{enumerate}
\end{thm}

The main technical difficulty for the proof of  theorem~\ref{lbpt3} consists of deducing 
an estimate of $\E \left(e^{M_T}-e^{M_T^n}\right)$ from an estimate of $\E\left(M_T-M_T^n\right)$.
In fact, the theorem can be deduced  from the following lemma.
\begin{lemme}\label{lbptlemme}
Assume that $X$ is an infinite activity L\'evy process with generating triplet $(\gamma,0,\nu)$ and there exists $\beta>1$ such that $\E e^{\beta M_T}<\infty$. Then for any $\epsilon>0$
\begin{eqnarray*}
\E \left(e^{M_T}-e^{M_T^n}\right)\leq C\left(\E\left(M_T-M_T^n\right)\right)^{\frac{\beta-1}{\beta}-\epsilon},
\end{eqnarray*}
where $C$ is a positive constant.
\end{lemme}

\begin{pf} \itshape lemma~\ref{lbptlemme}. \upshape
By the convexity of the exponential function, we have
\begin{eqnarray*}
e^{M_T}-e^{M_T^n}&\leq&\left(M_T-M_T^n\right)e^{M_T}.
\end{eqnarray*}
So, by H\"older's inequality,
\begin{eqnarray*}
\E \left(e^{M_T}-e^{M_T^n}\right)&\leq&\left(\E e^{\beta M_T}\right)^{\frac{1}{\beta}}\left(\E\left(M_T-M_T^n\right)^{\frac{\beta}{\beta-1}}\right)^{\frac{\beta-1}{\beta}}.
\end{eqnarray*}
Note that $\E e^{\beta M_T}<\infty$ implies that $\E M_T^{q}<\infty$ for any $q>0$. Let $\rho\in]0,1[$, we have
\begin{eqnarray*}
\E\left(M_T-M_T^n\right)^{\frac{\beta}{\beta-1}}&=&\E\left(M_T-M_T^n\right)^{\rho}\left(M_T-M_T^n\right)^{\frac{\beta}{\beta-1}-\rho}
\\&=&\E\left(M_T-M_T^n\right)^{\rho}\left(M_T-M_T^n\right)^{\frac{\beta(1-\rho)+\rho}{\beta-1}}
\\&\leq&\left(\E\left(M_T-M_T^n\right)\right)^{\rho}\left(\E\left(M_T-M_T^n\right)^{\frac{\beta(1-\rho)+\rho}{(\beta-1)(1-\rho)}}\right)^{1-\rho}.
\end{eqnarray*}
Hence, from the fact that $\lim_{n\rightarrow+\infty}\E\left(M_T-M_T^n\right)^{\frac{\beta(1-\rho)+\rho}{(\beta-1)(1-\rho)}}=0$, there exists a constant $C>0$ such that
\begin{eqnarray*}
\E \left(e^{M_T}-e^{M_T^n}\right)&\leq&C\left(\E\left(M_T-M_T^n\right)\right)^{\rho\frac{\beta-1}{\beta}}
\\&=&C\left(\E\left(M_T-M_T^n\right)\right)^{\frac{\beta-1}{\beta}-(1-\rho)\frac{\beta-1}{\beta}}.
\end{eqnarray*}
Then for any $\epsilon>0$, there exists a constant $C>0$ such that
\begin{eqnarray*}
\E \left(e^{M_T}-e^{M_T^n}\right)&\leq&C\left(\E\left(M_T-M_T^n\right)\right)^{\frac{\beta-1}{\beta}-\epsilon}.
\end{eqnarray*}
\end{pf}

When the L\'evy process driving the underlying asset has no  positive jumps, we get tighter estimates.\begin{prop}\label{lbpt4}
Let $X$ be a L\'evy process with generating triplet $(\gamma,\sigma^2,\nu)$. We assume that $X$ has no positive jump ($\nu(0,+\infty)=0$), that $\int_{-1\leq x<0}|x|\nu(dx)\\<\infty$ and that there exists $\beta>1$ such that $\E e^{\beta M_T}<\infty$. Then, the price of a discrete put lookback at the $k$-{th} fixing date and its continuous version at  time $k\Delta t$, satisfy
\begin{enumerate}
\item if $\sigma=0$
\begin{eqnarray*}
&&V_n\left(S_{+}\right)=V\left(S_{+}\right)+ O\left(\frac{1}{n}\right).
\end{eqnarray*}
\item if $\sigma>0$
\begin{eqnarray*}
&&V_n\left(S_{+}\right)=V\left(S_{+}\right)+  O\left(\frac{\log(n)}{\sqrt{n}}\right).
\end{eqnarray*}
\end{enumerate}
\end{prop}
Proposition~\ref{lbpt4} is based on the estimation of the moments of $M_T-M_T^n$, which can be
performed when there are no positive jumps.

\begin{lemme}\label{OprobMn4lemme}
Let $X$ be a L\'evy process with generating triplet $(\gamma,\sigma^2,\nu)$, satisfying $\int_{|x|\leq1}|x|\nu(dx)<\infty$. We suppose that $X$ has no positive jumps, then for any $\beta>1$, we have
\begin{enumerate}
 \item if $\sigma=0$,
       \begin{eqnarray*}
         \E\left(M_T-M_T^n\right)^{\beta}= O\left(\frac{1}{n^{\beta}}\right).
       \end{eqnarray*}
 \item if $\sigma>0$,
       \begin{eqnarray*}
         \E\left(M_T-M_T^n\right)^{\beta}= O\left(\left(\frac{\log(n)}{\sqrt{n}}\right)^{\beta}\right).
       \end{eqnarray*}       
\end{enumerate}
\end{lemme}

\begin{pf} \itshape lemma~\ref{OprobMn4lemme}. \upshape
We have
\begin{eqnarray*}
M_T-M_T^n&=&\sup_{0\leq s\leq T}X_s-\max_{0\leq k\leq n}X_{\frac{kT}{n}}
\\&=&\max_{1\leq k\leq n}\sup_{\frac{(k-1)T}{n}\leq s\leq \frac{kT}{n}}X_s-\max_{0\leq k\leq n}X_{\frac{kT}{n}}
\\&\leq&\max_{1\leq k\leq n}\sup_{\frac{(k-1)T}{n}\leq s\leq \frac{kT}{n}}X_s-\max_{1\leq k\leq n}X_{\frac{(k-1)T}{n}}
\\&\leq&\max_{1\leq k\leq n}\left(
    \sup_{\frac{(k-1)T}{n}\leq s\leq \frac{kT}{n}}X_s-X_{\frac{(k-1)T}{n}}\right),
\end{eqnarray*}
where the random variables 
$\left(\sup_{\frac{(k-1)T}{n}\leq s\leq \frac{kT}{n}}X_s-X_{\frac{(k-1)T}{n}}\right)_{1\leq k\leq n}$
 are i.i.d., with the same distribution as $\sup_{0\leq s\leq \frac{T}{n}}X_s$. But, since  $X$ has no positive jumps, we have (see \eqref{levyactiviteinfinievf})
\begin{eqnarray*}
\sup_{0\leq s\leq \frac{T}{n}}X_s&\leq&\sup_{0\leq s\leq \frac{T}{n}}\left(\gamma_0 s+\sigma B_s\right)
\\&\leq&\frac{|\gamma_0|T}{n}+\sigma\sup_{0\leq s\leq \frac{T}{n}}B_s.
\end{eqnarray*}
We can easily deduce the first result of the lemma ($\sigma=0$). In the case $\sigma>0$, we have
\begin{eqnarray*}
\sup_{0\leq s\leq \frac{T}{n}}X_s&\leq&\frac{1}{\sqrt{n}}\left(\frac{|\gamma_0|T}{\sqrt{n}}+\sigma\sqrt{n}\sup_{0\leq s\leq \frac{T}{n}}B_s\right)
\\&\leq&\frac{1}{\sqrt{n}}\left(|\gamma_0|T+\sigma\sqrt{n}\sup_{0\leq s\leq \frac{T}{n}}B_s\right)
\\&=^d&\frac{1}{\sqrt{n}}\left(|\gamma_0|T+\sigma\sup_{0\leq s\leq T}B_s\right).
\end{eqnarray*}
Let $\left(V_k\right)_{1\leq k\leq n}$ be i.i.d. r.v. with the same distribution as $|\gamma_0|T+\sigma\sup_{0\leq s\leq T}B_s$. Then we have
\begin{eqnarray*}
\E\left(M_T-M_T^n\right)^{\beta}&\leq&\left(\frac{1}{\sqrt{n}}\right)^{\beta}\E\max_{1\leq k\leq n}V_k^{\beta}.
\end{eqnarray*}
Let $g$ be the function defined as follows 
\begin{equation*}
g(x)=\left(\log(x)\right)^{\beta}, \quad x>1.
\end{equation*}
The function $g$ is concave and non-decreasing on 
the set $[e^{\beta-1},+\infty)$. So we have
\begin{eqnarray*}
\E\sup_{1\leq k\leq n}V_k^{\beta}&=&\E\sup_{1\leq k\leq n}g\left(e^{V_k}\right)\\
&=&\E g\left(\sup_{1\leq k\leq n}e^{V_k}\right), \ \mbox{because g is non-decreasing}\\
&\leq&\E g\left(\sup_{1\leq k\leq n}e^{\max\left( V_k,\beta-1\right)}\right), \ \mbox{because g is non-decreasing}\\
&\leq&g\left(\E\sup_{1\leq k\leq n}e^{\max\left( V_k,\beta-1\right)}\right), \ \mbox{by Jensen}\\
&\leq&g\left(\E\sum_{k=1}^ne^{\max\left( V_k,\beta-1\right)}\right), \ \mbox{because g is non-decreasing}\\
&\leq&g\left(n\E e^{\max\left( V_1,\beta-1\right)}\right).
\end{eqnarray*}
Note that we have $\E e^{\max\left(V_1,\beta-1\right)}<\infty$. Hence the second result of the lemma.
\end{pf}

\begin{pf} \itshape proposition~\ref{lbpt4}. \upshape 
To prove proposition~\ref{lbpt4}, we need to show that
\begin{eqnarray*}
\E\left(e^{M_T}-e^{M_T^n}\right)=\left\{\begin{aligned}
                                     O\left(\frac{1}{n}\right)\ \ \ \ \ \ \ if \ \sigma=0\\
                                     O\left(\frac{\log(n)}{\sqrt{n}}\right)\ if \ \sigma>0                                     
                                   \end{aligned}\right.
\end{eqnarray*}
But by the convexity of the exponential function, we have
\begin{eqnarray*}
e^{M_T}-e^{M_T^n}\leq e^{M_T}\left(M_T-M_T^n\right).
\end{eqnarray*}
So using H\"older's inequality, we get
\begin{eqnarray*}
\E\left(e^{M_T}-e^{M_T^n}\right)&\leq&\left(\E e^{\beta M_T}\right)^{\frac{1}{\beta}}\left(\E\left(M_T-M_T^n\right)^{\frac{\beta}{\beta-1}}\right)^{\frac{\beta-1}{\beta}}.
\end{eqnarray*}
We conclude by lemma~\ref{OprobMn4lemme}.
\end{pf}

Results for hindsight options are similar to those for lookback options. This is simply due to the relations between lookback and hindsight options.

\end{document}